\documentclass[english,aps,prc,twocolumn,showpacs,amsmath,amssymb,preprintnumbers, superscriptaddress]{revtex4}
\usepackage[T1]{fontenc}
\usepackage{babel}
\usepackage{amsmath}
\usepackage{amssymb}
\usepackage{graphicx}
\usepackage{color}
\usepackage{comment}
\usepackage[utf8]{inputenc}

\usepackage{times}
\usepackage{hyperref}
\hypersetup{
  colorlinks=true,        
  linkcolor=blue,         
  citecolor=cyan,         
}

\definecolor{orange}{rgb}{1,0.5,0}
\definecolor{darkgreen}{rgb}{0,0.7,0}

\newcommand{\ie}{i.e.,~}
\newcommand{\eg}{e.g.,~}

\newcommand{\al}{\alpha}

\newcommand{\de}{\delta}
\newcommand{\De}{\Delta}

\newcommand{\ka}{\kappa}
\newcommand{\la}{\lambda}

\newcommand{\ba}{\begin{array}}
\newcommand{\ea}{\end{array}}
\newcommand{\bea}{\begin{eqnarray}}
\newcommand{\eea}{\end{eqnarray}}
\newcommand{\bi}{\begin{itemize}}  
\newcommand{\ei}{\end{itemize}}
\newcommand{\ben}{\begin{enumerate}} 
\newcommand{\een}{\end{enumerate}}
\newcommand{\bc}{\begin{center}}
\newcommand{\ec}{\end{center}}

\newcommand{\save}[1]{}
\usepackage[normalem]{ulem}  

\newcommand{\MeV}{{\rm MeV}}
\newcommand{\ms}{\,{\rm ms}}

\begin{document}

\title{On the importance of viscous dissipation and heat conduction in
  binary neutron-star mergers}

\author{Mark G. Alford} 
\affiliation{Physics Department, Washington University, St.~Louis,
  MO~63130, USA}
\author{Luke Bovard} \affiliation{Institut f\"ur Theoretische
  Physik, Max-von-Laue-Strasse 1, 60438 Frankfurt, Germany}
\author{Matthias Hanauske} \affiliation{Institut f\"ur Theoretische
  Physik, Max-von-Laue-Strasse 1, 60438 Frankfurt, Germany}
\affiliation{Frankfurt Institute for Advanced Studies,
  Ruth-Moufang-Strasse 1, 60438 Frankfurt, Germany}
\author{Luciano Rezzolla}
\affiliation{Institut f\"ur Theoretische Physik, Max-von-Laue-Strasse 1,
  60438 Frankfurt, Germany}
\affiliation{Frankfurt Institute for Advanced Studies,
  Ruth-Moufang-Strasse 1, 60438 Frankfurt, Germany}
\author{Kai Schwenzer}
\affiliation{Theoretical Astrophysics (IAAT), Eberhard Karls University
  of T\"ubingen, T\"ubingen 72076, Germany}
\affiliation{Department of Astronomy and Space Sciences, Istanbul
  University, Beyaz\i t, 34119, Istanbul, Turkey}
 
\begin{abstract}
Inferring the properties of dense matter is one of the most exciting
prospects from the measurement of gravitational waves from neutron star
mergers. However, it will require reliable numerical simulations that
incorporate viscous dissipation and energy transport if these can play a
significant role within the survival time of the post-merger object. We
calculate timescales for typical forms of dissipation and find that
thermal transport and shear viscosity will not be important unless
neutrino trapping occurs, which requires temperatures above about 10\,MeV
and gradients over lengthscales of 0.1\,km or less. On the other hand, if
direct-Urca processes remain suppressed, leaving modified-Urca processes
to establish flavor equilibrium, then bulk viscous dissipation could
provide significant damping to density oscillations observed right
after the merger. When comparing with data from a state-of-the-art merger
simulation we find that the bulk viscosity takes values close to its
resonant maximum in a typical neutron-star merger, motivating a more
careful assessment of the role of bulk viscous dissipation in the
gravitational-wave signal from merging neutron stars.
\end{abstract}


\maketitle

\smallskip
\noindent {\em Introduction.}  LIGO's first direct detection of
gravitational waves from the merger of black holes~\cite{Abbott:2016blz}
has improved enormously the prospects for the detection of gravitational
waves from neutron-star mergers in the near future. Because neutron stars
are extended structures of ultra-dense matter, such observations could
contain valuable information about the properties of matter at extreme
density and temperature, and the central engine of short gamma-ray bursts
(see~\cite{Baiotti:2016qnr, Paschalidis2016} for reviews). With a few
exceptions~\cite{Duez:2004nf, Shibata:2017b} current numerical
simulations of neutron-star mergers neglect the transport properties of
the material on the assumption that they will operate on timescales much
larger than the dynamical ones of the binary~\cite{Bildsten92}. In this
study we revisit this assumption by exploring the impact of shear/bulk
viscosity and thermal transport immediately after the merger and by
exploiting the results of numerical relativity. These simulations have
seen enormous progress in recent years~\cite{Shibata99d, Baiotti08,
  Anderson2007, Liu:2008xy, Bernuzzi2011} and have found that if the
total mass of the system is not too large, then the post-merger object is
metastable to gravitational collapse over a timescale of tens of
milliseconds. The inner region of this object, which is about 10\,km
across, can reach densities of several times nuclear-matter saturation
(number) density $n_0\approx 0.16\,{\rm fm}^{-3}$ and temperatures of
tens of MeV.

The details of the complicated post-merger phase depend on the mass of
the system, the equation of state (EOS), and the strength of the magnetic
fields that develop after the merger~\cite{Baiotti:2016qnr,
  Paschalidis2016}. Quite generically, however, unless it collapses
promptly to a black hole \cite{Baiotti08}, the binary-merger product will
oscillate in modes that leave a clear imprint on the gravitational-wave
signal~\cite{Bauswein2011, Stergioulas2011b, Takami2014, Bernuzzi2015a,
  Rezzolla2016}. In such simulations, the large scale motion is damped by
gravitational-wave emission over tens of milliseconds. This sets the
timescale over which dissipation or transport would have to operate in
order to influence the details and duration of the gravitational-wave
signal or the emission of neutrinos. Our goal in this \emph{Letter} is to
provide simple estimates of the likely effects on the merger product of
the transport properties of the high-density matter. 
Transport properties vary greatly between
different phases, offering the possibility of using data from mergers to
probe the phase structure of dense matter, 
potentially including exotic phases such as quark matter.
In particular, we
provide estimates of the thermal transport and dissipation timescales
over which these processes have a noticeable effect on the system, as a
function of relevant properties of the material such as rest-mass density
and temperature. 

\smallskip
\noindent {\em Thermal equilibration.}  To see whether heat diffusion is
a significant effect, we first estimate the thermal equilibration time.
For this, consider a region of size $z_{\rm typ}$ that is hotter than its
surroundings by a temperature difference $\De T$. For a material with
specific heat per unit volume $c_V$ and thermal conductivity $\kappa$,
this region has an extra thermal energy $E_{\rm th} \!\approx \!(\pi/6)
c_V z_{\rm typ}^3 \De T $ and (assuming a smooth temperature distribution
so that the thermal gradient is $\De T/z_{\rm typ}$) heat is conducted
out of the region at a rate $W_{\rm th} \!\approx \!\pi \kappa \De T
z_{\rm typ}$. The thermal equilibration time, namely the time needed to
conduct away a significant fraction of the extra thermal energy, is
$\tau_\kappa \!\equiv\! E_{\rm th}/W_{\rm th} = c_V z_{\rm
  typ}^2/(6\kappa)$.
Hence, to evaluate $\tau_\kappa$ we need estimates of the specific heat
and thermal conductivity of beta-equilibrated nuclear matter. The
specific heat is dominated by neutrons, which have the largest phase
space of low-energy excitations, giving $c_V\approx 1.0\, m_n^*
n_n^{1/3}\, T$, assuming a Fermi liquid of neutron density $n_n$ with
Landau effective mass $m_n^*$ \cite{Levenfish:1994}.

For the thermal conductivity, we recall that in kinetic theory particles
of number density $n_i$, typical speed $v_i$, and mean free path (mfp)
$\la_i$, contribute to the thermal conductivity as $\kappa \propto \sum_i
\kappa_i \propto \sum_i n_i v_i \la_i$, so that $\kappa$ is effectively
dominated by particles with the optimal combination of high density and
long mfp. The neutrons, though numerous, are strongly interacting and
have a very short mfp. Thermal conductivity is therefore dominated by
electrons or neutrinos. At low temperatures, \ie below a few MeV, the
neutrino mfp becomes longer than the size of the merger
region~\cite{Baiotti:2016qnr, Reddy:1997yr}, so neutrinos escape and
thermal conductivity is dominated by electrons which scatter via exchange
of Landau-damped transverse photons. The thermal conductivity is then
temperature-independent $\ka_e \approx 1.5\, n_e^{2/3}/\al$ [see
  Eq.~(40) of Ref.~\cite{Shternin:2007ee}], where $n_e$ is the electron
number density and $\al\approx1/137$ is the fine-structure
constant. This yields the thermal equilibration time in the electron
dominated regime
\begin{align} 
\tau_\ka^{(e)} & = 4.7\times10^8\,{\rm s} \nonumber \\
& \left(\dfrac{0.1}{x_p}\right)^{\tfrac{2}{3}} 
\left(\dfrac{m_n^*}{0.8\,m_n}\right) 
\left(\dfrac{n_0}{n_n}\right)^{\tfrac{1}{3}}   \left(\dfrac{z_{\rm
    typ}}{1\,{\rm km}}\right)^{2}  \left(\dfrac{T}{1\,\MeV}\right)
\,, \nonumber
\end{align}
where $x_p \equiv n_e/n_n$ is the proton fraction. Clearly, this
timescale is far too large to have an impact on the $\sim \!10$\,ms
timescale of post-merger processes~\cite{Baiotti:2016qnr}.

At temperatures $T\!\gtrsim\!10$ MeV, neutrinos become trapped for
nucleon density $n \gtrsim n_0$, since the neutrino mfp, which at high
density depends strongly on in-medium corrections
\cite{Reddy:1997yr,Roberts:2016mwj}, becomes smaller than the
star. Electron neutrinos form a degenerate Fermi gas with a Fermi
momentum $p_{F,\nu}$ of about half that of the electrons. Their mfp is
longer than that of the electrons, so they dominate the thermal
conductivity~\cite{Goodwin:1982hy}, which is given by $\kappa_\nu\approx
0.33\, n_{\nu}^{2/3}/(G_F^{2}(m_{n}^{*})^2 n_{e}^{1/3} T)$, where $G_F
\equiv 1/(293\,\MeV)^2$ is the Fermi coupling. This yields the timescale
for thermal transport via neutrinos
\begin{align} 
& \tau_\kappa^{(\nu)} \approx  0.7\,{\rm s} \nonumber \\
& \times\left(\dfrac{0.1}{x_p}\right)^{\!\tfrac{1}{3}}\!\!\left(
\dfrac{m_n^*}{0.8\,m_n}\right)^{\!3}
\!\!\left(\dfrac{\mu_{e}}{2\,\mu_{\nu}}\right)^{\!2} 
\!\!\left(\dfrac{z_{{\rm typ}}}{1\,{\rm km}}\right)^{\!2} 
\!\!\left(\dfrac{T}{10\,{\rm MeV}}\right)^{\!2} .
\label{thermal-conductivity-scale}
\end{align}

In summary, for neutrino-driven thermal transport to be important on a
timescale of tens of milliseconds, there would e.g. have to be thermal
gradients (\eg from turbulence) on lengthscales of the order
0.1\,km. Moreover, heat transport into cooler regions should manifest
itself even more quickly.

\smallskip
\noindent {\em Shear dissipation.} We next estimate the timescale on
which shear viscosity plays a role. For this, consider a fluid of
rest-mass density $\rho$ flowing in the $x$ direction at velocity
$v_{x}$, having kinetic energy per unit volume $E_{\rm
  kin}=\frac{1}{2}\rho v_{x}^{2}$. If the fluid has shear viscosity
$\eta$, then the energy dissipated per unit time and unit volume is
$W_{\rm shear} \approx \eta (dv_{x}/dz)^{2}$, so that the time needed for
shear viscosity to dissipate a significant fraction of the kinetic energy
is $\tau_{\eta} \equiv E_{\rm kin}/W_{\rm shear}$. To estimate
$\tau_{\eta}$, we assume that the flow is fairly uniform, with the
velocity varying by a factor of order unity over a distance $z_{{\rm
    typ}}$ in the $z$ direction, so $dv_{x}/dz \approx v_{x} / z_{\rm
  typ}$ which gives $\tau_{\eta} \approx \rho\,z_{{\rm
    typ}}^{2}/(2\eta)$.

In the low-temperature, electron-dominated regime (\ie $T\lesssim
10\,\MeV$), using the dominant transverse contribution from
Ref.~\cite{Shternin:2008es} [see Eq.~(2.4) in Ref.~\cite{Manuel:2012rd}]
with the damping scale $q_t^2 \!\equiv\! 4 \al p_{F,e}^2/\pi$, we find
$\eta^{(e)} \approx 0.2 \, n_e^{14/9}/(\al^{5/3}\, T^{5/3})$, so
\begin{equation}
\tau_\eta^{(e)} \approx 1.6\!\times \!10^8\, {\rm s} \left( \dfrac{z_{\rm
    typ}}{1\,{\rm km}}\right)^{\! 2} \!\!  \left(
\!\dfrac{T}{1\,\MeV}\!\right)^{\!\tfrac{5}{3}} \!\! \left(
\!\dfrac{n_0}{n_B}\!\right)^{\!\tfrac{5}{9}} \!\!  \left(
\!\dfrac{0.1}{x_p}\!\right)^{\!\tfrac{14}{9}} \,, 
\end{equation}
where $n_B$ is the baryon number density of nuclear matter. As a result,
when electrons dominate, it would take years for the shear viscosity to
significantly impact the flow. However, in the high-temperature,
neutrino-dominated regime (\ie for $T\gtrsim 10\,\MeV$) neutrinos produce
a much larger shear viscosity $\eta^{(\nu)} \approx 0.46\,
n_{\nu}^{4/3}/\left(G_F^{2} (m_{n}^{*})^2
n_{e}^{1/3}T^{2}\right)$~\cite{Goodwin:1982hy}, which yields
\begin{equation}
\tau_{\eta}^{(\nu)} \approx 54\,{\rm s}\, \left(\dfrac{0.1}{x_p}\right)
\!\left( \!\dfrac{m_n^*}{0.8\,m_n}\!\right)^{\! 2} \!\!\left(
\!\dfrac{\mu_{e}}{2\,\mu_{\nu}}\!\right)^{\! 4} \!\! \left(\dfrac{z_{{\rm
      typ}}}{1\,{\rm km}}\right)^{\! 2} \!\! \left( \!\dfrac{T}{10\,{\rm
    MeV}}\!\right)^{\! 2}\,,
\label{shear-viscosity-scale}
\end{equation}
Interestingly, this result depends only indirectly on the density, via
the proton fraction $x_p$ and the ratio of lepton chemical potentials
$\mu_{e}, \mu_{\nu}$. In summary, neutrino shear viscosity could play an
important role, \ie $\tau_{\eta}^{(\nu)}$ could be in the millisecond
range, if the neutrino density is anomalously high or if there are flows
that experience shear over short distances, $z_{\rm typ} \sim 0.01\,{\rm
  km}$, for example, due to turbulence or high-order non-axisymmetric
instabilities~\cite{Kiuchi:2015sga, East2016, Radice2016a, Lehner2016a}.

\smallskip
\noindent {\em Bulk viscosity.} Next, we study the impact of bulk
viscosity, which characterizes the degree to which there is production of
heat when a material is compressed or rarefied. We will consider an
``averaged'' bulk viscosity $\bar\zeta$ in response to a periodic
compression-rarefaction cycle. In nuclear matter under compression on
millisecond timescales, dissipation arises because the rate of beta
equilibration of the proton fraction via Urca processes occurs on the
same timescale, allowing the proton fraction to fall out of phase with
the applied pressure. As long as the oscillations in the binary-merger
product are roughly periodic, we expect that the dissipation induced by
pressure variations occurring on a timescale $t_{\rm dens}$ can be
estimated by using the bulk viscosity evaluated at frequency $f=1/t_{\rm
  dens}$. For the interactions of interest here, the bulk viscosity is
largest when the internal equilibration rate matches the frequency of the
oscillation. Furthermore, because the equilibration rate is sensitive to
the temperature, the bulk viscosity shows a resonant maximum as a
function of temperature (see, \eg Fig.~7 in~\cite{Alford:2010gw}). For
oscillations with a timescale $t_{\rm dens}$, the resonant maximum value
is \cite{Alford:2010gw}
\begin{align}
\bar\zeta_{{\rm max}} 
\equiv Y_{\zeta}\, \bar n \, t_{\rm dens}\,, & &
 Y_{\zeta} \equiv C^{2}/(4\pi B \bar n)\,,
\label{bulk:max}
\end{align}
where $B \!\equiv\! -\left(1/\bar{n}\right) \left.\left(\partial
\delta\mu/\partial x_p\right)\right|_{n}$ and
$C\!\equiv\!\bar{n}\left.\left(\partial\delta\mu/\partial n
\right)\right|_{x_p}$ are the nuclear susceptibilities with respect to
baryon density and proton fraction, where the chemical potential
$\delta\mu\!\equiv\!\mu_n\! - \!\mu_p\! - \!\mu_e$ characterises the
degree to which the system is out of beta equilibrium. This maximum value
$\bar\zeta_{{\rm max}}$ depends only on properties of the EOS, that is,
it is {\em independent} of the flavor re-equilibration rate. Changing the
re-equilibration rate moves the curve in Fig.~7 in \cite{Alford:2010gw}
``horizontally'', changing the temperature at which the maximum value is
attained.

\begin{figure}
\center
\includegraphics[width=\columnwidth]{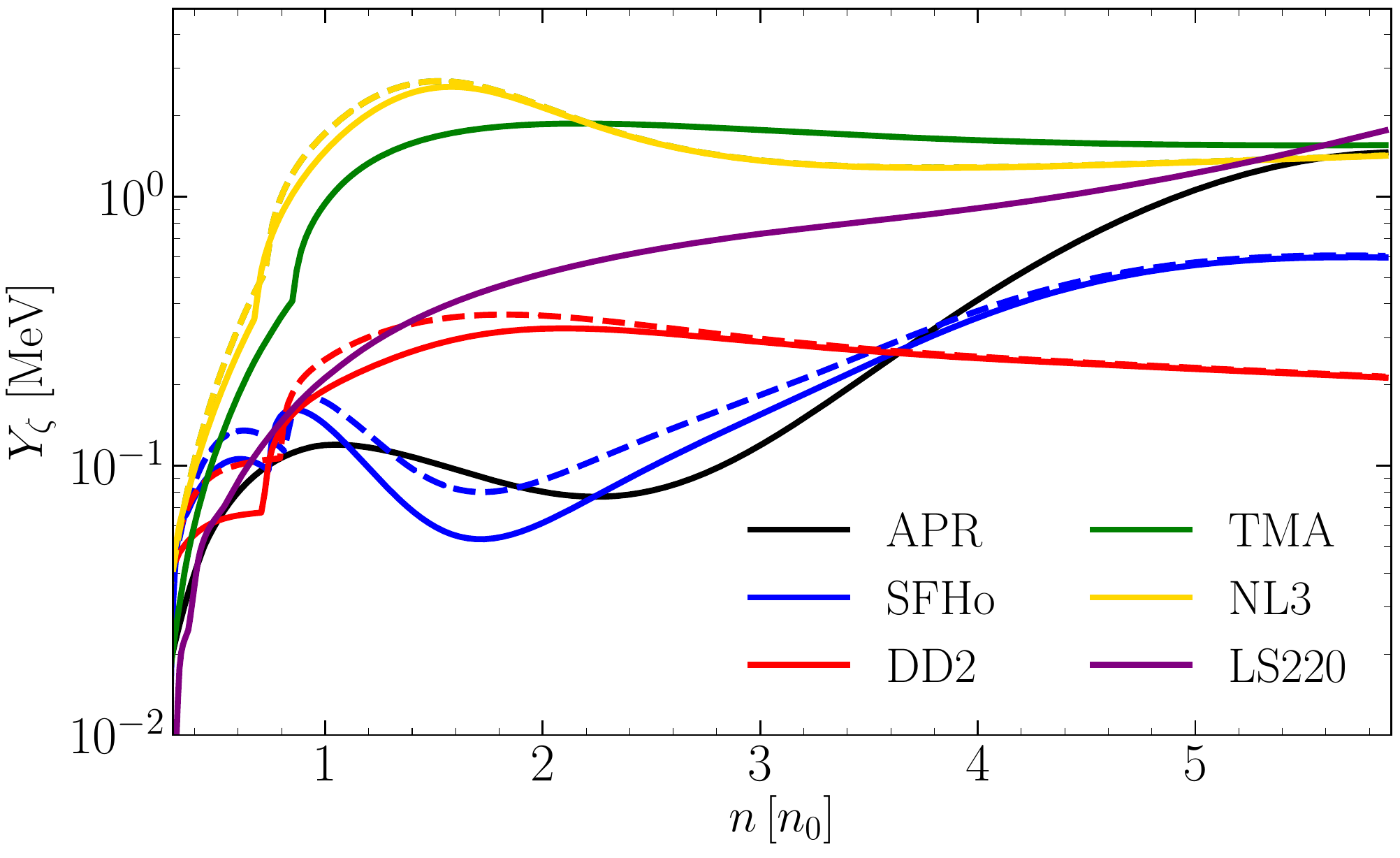}
\caption{Density dependence of the maximum-bulk-viscosity prefactor
  $Y_\zeta$ [see Eq.~(\ref{bulk:max})] for various EOSs. Solid lines are
  for cold matter ($T\!=\!0.1\,\MeV$) while dashed lines are for hot matter
  ($T\!=\!10\,\MeV$). For LS220 we only give a single curve at $T\!=\!1\, \MeV$, due to numerical issues in the EOS table.
\label{fig:bulk-prefactor}
}
\end{figure}

We note that the maximum bulk viscosity is a monotonically increasing
function of number density and Fig.~\ref{fig:bulk-prefactor} shows the
prefactor $Y_{\zeta}$ for nuclear matter obeying various EOSs, all of
which can sustain a $2\, M_\odot$ neutron star~\cite{Demorest:2010bx,
  Antoniadis2013}. Whereas APR~\cite{Akmal:1998cf} is a cold EOS and is
included here for comparison, for all the others we use ``hot'' EOSs
calculated using a model of nuclei and interacting nucleons in
statistical equilibrium~\cite{Hempel:2009mc}. In addition to the LS220 
\cite{Lattimer:1991nc} (typical radius 13\,km), 
used for the simulations below, these EOSs range
from the moderately soft SHFo~\cite{Steiner:2012rk}, yielding a typical
radius of 11\,km, through the increasingly stiff DD2
(13\,km)~\cite{Steiner:2012rk, Fischer:2013eka} and TMA
(14\,km)~\cite{Hempel:2009mc}, to the extremely stiff NL3, with a typical
radius of 15 km and a maximum mass of nearly $3\, M_\odot$. For APR we
use a standard quadratic parametrization in terms of the symmetry energy
\cite{Alford:2010gw} to compute the susceptibilities.

We now show that the temperature $T_{\zeta{\rm max}}$ at which bulk
viscosity reaches its resonant maximum for $\sim \!1$\,kHz density
oscillations (assuming flavor equilibrates via nuclear modified-Urca
"nmU" processes) is in the range of temperatures typically found in
neutron-star mergers. We recall that for small-amplitude oscillations
$T_{\zeta{\rm max}}=( 2\pi f/(\tilde{\Gamma}B)
)^{1/\de}$~\cite{Alford:2010gw}, where $\tilde{\Gamma}$ is the prefactor
in the equilibration rate, $\Gamma=\tilde\Gamma T^\delta \delta\mu$. For
modified-Urca processes, $\delta\!=\!6$, so $1/\delta$ is small, making
$T_{\zeta{\rm max}}$ insensitive to details of the EOS. As a result, over
the relevant frequency range, \ie from a few tenths to several kHz, we
find for "nmU" processes
\begin{equation}
T_{\zeta{\rm max}}^{{\rm nmU}}\approx4-7\,{\rm
  MeV}\approx5-8\times10^{10}\,{\rm K}\,,
\label{bulk:Tmax-value}
\end{equation}
which is well within the range of temperatures expected for dense matter in the
post-merger~\cite{Baiotti:2016qnr, Dietrich2016, Hanauske2016}.

\begin{figure}
\center
\includegraphics[width=\columnwidth]{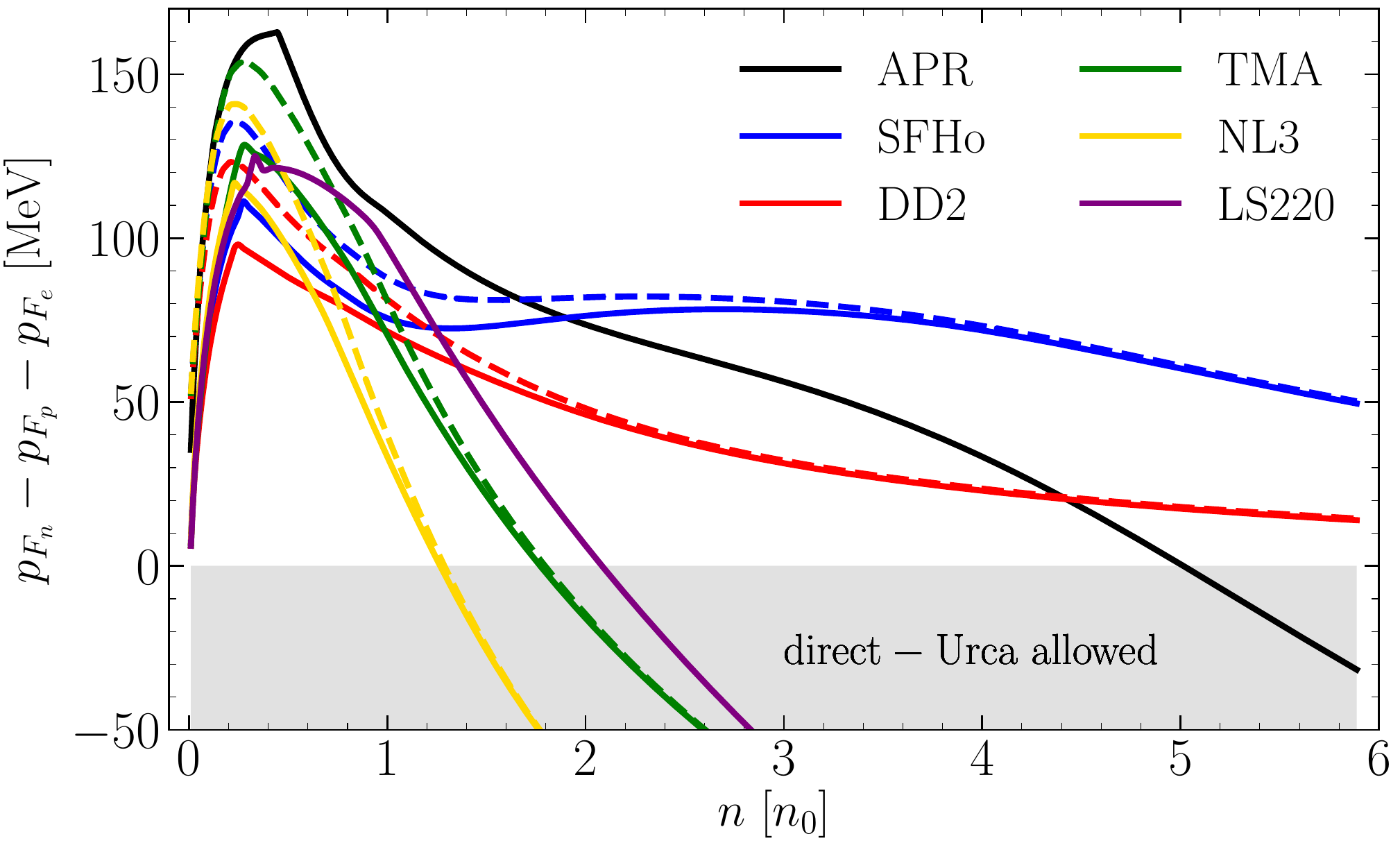}
\caption{Momentum difference relevant to the direct-Urca channel as a
  function of density, for the EOSs shown in
  Fig.~\ref{fig:bulk-prefactor}. For negative values, direct-Urca
  processes are allowed (grey-shaded area).}
\label{fig:kin-constraint}
\end{figure}

It should be noted that flavor re-equilibration might instead occur via
direct-Urca reactions, which are orders of magnitude faster than
modified-Urca processes, giving much lower bulk viscosities at $T\sim
5\,\MeV$, since the resonant maximum of bulk viscosity would have moved
to lower temperatures (see Fig.~7 in Ref.~\cite{Alford:2010gw}). In
neutrino-transparent matter at $T\!=\!0$, direct-Urca processes are allowed
when $\Delta p_F \!\equiv\!  p_{F,n}\!-\!p_{F,p}\!-\!p_{F,e} \!< \!0$. In
Fig.~\ref{fig:kin-constraint} we plot this kinematic constraint as a
function of density for the same EOSs in
Fig.~\ref{fig:bulk-prefactor}. For softer EOSs (\eg SFHo, DD2)
direct-Urca processes are never possible at $T=0$; however, for APR the
direct-Urca channel opens at $n> 5\,n_0$ and for even stiffer EOSs (\eg
NL3, TMA) it already opens below twice saturation density. These
considerations suggest that the amount of bulk-viscous damping will be a
sensitive indicator of whether the EOS allows direct Urca processes at
the densities and temperatures prevalent in neutron star mergers. To draw
a more precise connection to the EOS will require calculations of the
beta equilibration rate that incorporate the effects of temperature,
strong interactions, and the gradual opening of phase space above the
direct Urca threshold.  The rest of our analysis will focus on showing
that if the direct-Urca channel is not open, then bulk viscosity can be
expected to have significant effects on the evolution of the post-merger
object.

We now estimate the dissipation time for compression oscillations. The
energy density for a baryon number-density oscillation of amplitude $\De
n$ around average density $\bar n$ is ${\cal E}_{\rm comp}\approx K \bar
n ( \De n/\bar n )^{2}/18$ \cite{Klahn:2006ir}, where $K$ is the nuclear
compressibility at that density. If the compression varies on a timescale
$t_{\rm dens}$, then, in a material with bulk viscosity $\bar\zeta$, the
dissipated power per unit volume is~\cite{Sawyer:1989dp} $\left( d{\cal
  E}/dt \right)_{\rm bulk} \approx 2 \pi^2 \bar\zeta \left(\Delta n/\bar
n \right)^2 /t_{\rm dens}^2$. Hence, the time required for bulk viscosity
to have a significant impact on the oscillations of the system is
\begin{equation}
\tau_\zeta \equiv {\cal E}_{\rm comp}/\left(d{\cal E}/dt\right)_{\rm
  bulk} \approx K \bar{n}\,t_{\rm dens}^2/(36 \pi^2\,\bar\zeta) \,.
\label{bulk:tau}
\end{equation}

Expecting bulk viscosity to reach its maximum value $\bar\zeta_{{\rm
    max}}$ [see Eq.~(\ref{bulk:max})] at typical neutron-star merger
temperatures [see Eq.~(\ref{bulk:Tmax-value})], we can use
Eq.~(\ref{bulk:max}) in (\ref{bulk:tau}) to find that the minimum
timescale for bulk viscosity to impact the oscillations is
\begin{equation}
\tau^{\rm min}_{\zeta} 
\approx 3 \,{\rm ms}\,\left( \dfrac{t_{\rm dens}}{1 \, {\rm ms}}\right) \,
\left(\dfrac{K}{250\,\MeV}\right)
\left(\dfrac{0.25\,{\rm MeV}}{Y_\zeta}\right)\,.
\label{bulk:tau-min}
\end{equation}

\begin{figure}
\center
\includegraphics[width=\columnwidth]{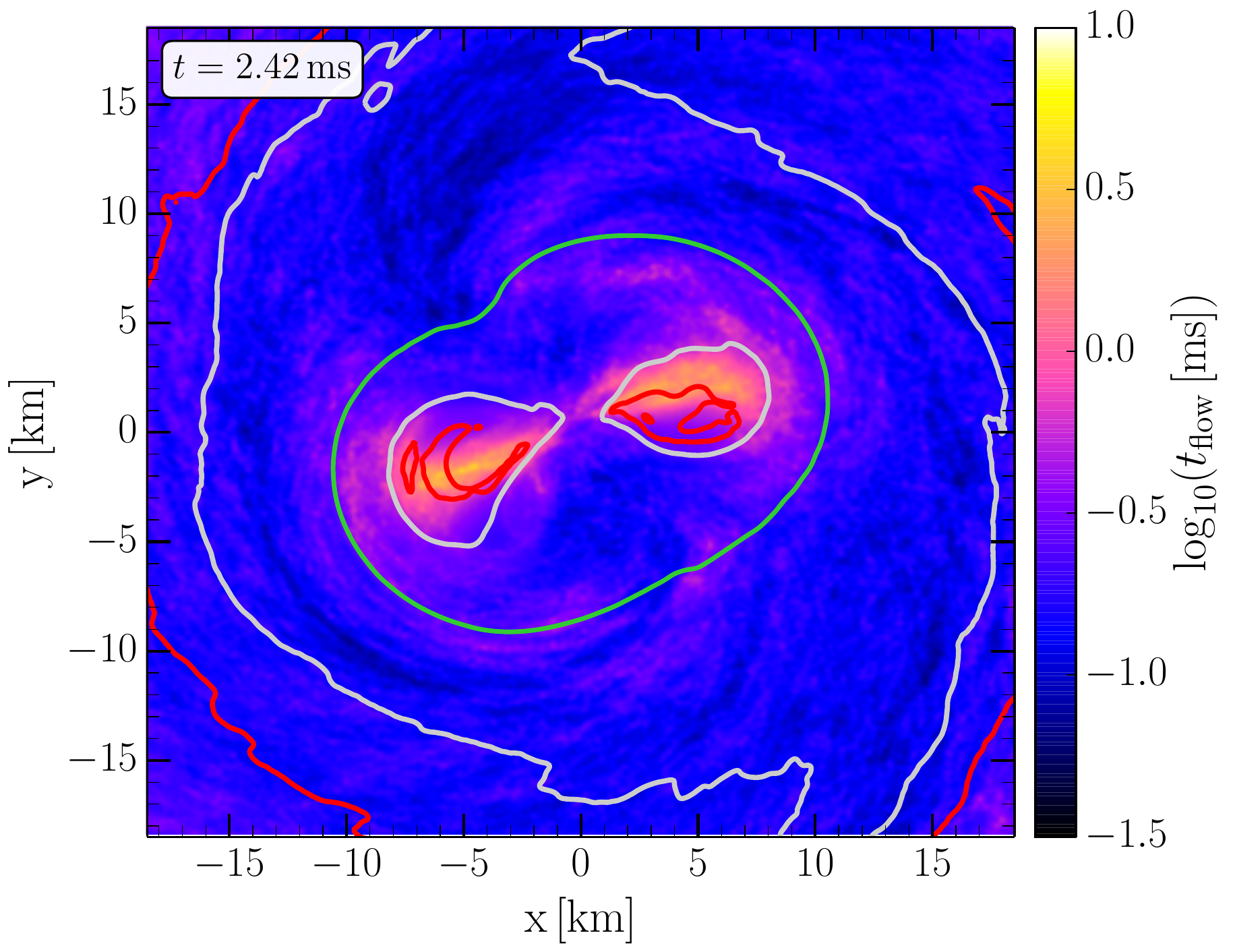}
\caption{The flow timescale $t_{\rm flow}$ obtained from a
  numerical-relativity simulation of two $1.35\,M_{\odot}$ neutron
  stars~\cite{Hanauske:2016gia}. The red (4 MeV) and gray (7 MeV)
  contours show the boundaries of the temperature range in which the bulk
  viscosity roughly takes its maximum value, while the green contour
  shows the inner region where the rest-mass density exceeds nuclear
  saturation density.}
\label{fig:t_typ}
\end{figure}

In essence, Eq.~(\ref{bulk:tau-min}) reveals that under conditions
of maximum bulk viscosity, the
damping timescale is a few times larger than the typical timescale
$t_{\rm dens}$ of density variations. 

Since a strong emission of gravitational waves occurs from the
high-density region of the star during the first $\sim 5$ milliseconds
after the merger, when characteristic frequencies $f_1$ and $f_3$ appear
in the gravitational-wave spectrum~\cite{Takami2014, Takami2015,
  Rezzolla2016}, bulk viscous damping is most likely to have observable
consequences if, during that early time, there are density oscillations
occurring on a millisecond timescale in parts of the high-density region
where the bulk viscosity is maximal (\ie for $T \sim 4\!-\!7\,{\rm MeV}$).

\begin{figure}
\center 
\includegraphics[width=0.45\textwidth]{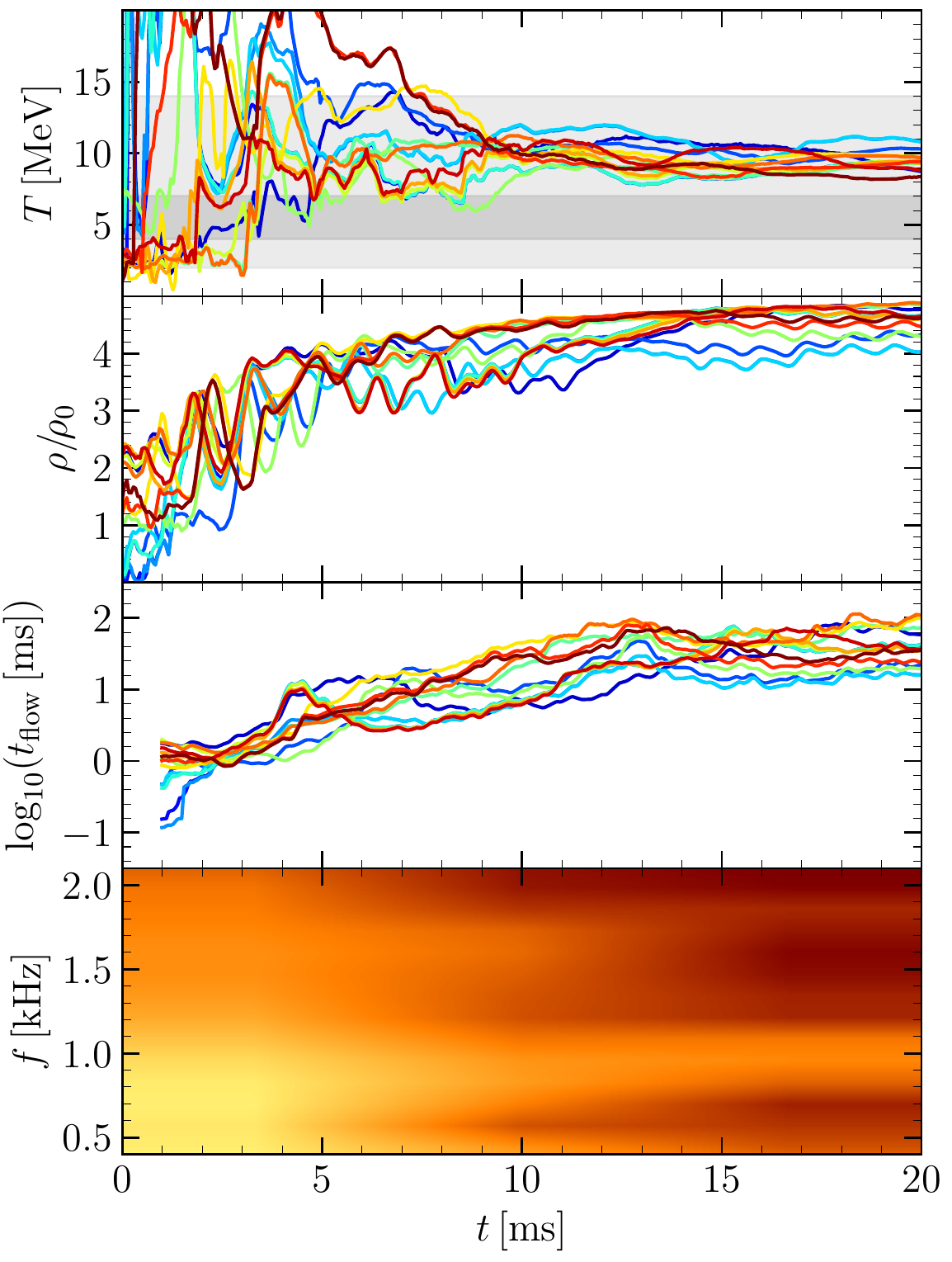}
\caption{Co-moving time variation of physical properties of post-merger
  material from selected tracers in the same merger as shown in
  Fig.~\ref{fig:t_typ}. Top panel: temperature [the shaded regions are
    where bulk viscosity is large, see
    Eq.~\eqref{bulk:Tmax-value}]. Second panel: rest-mass density. Third
  panel: flow timescale $t_{\rm flow}$. Bottom panel: spectrogram
  averaging the rest-mass density evolutions in the second panel.}
\label{fig:tracers}
\end{figure}

To test whether such conditions are met, we show in Figs.~\ref{fig:t_typ}
and \ref{fig:tracers} results from a state-of-the-art simulation of a
merger of two $M=2\times 1.35\,M_{\odot}$ stars using the LS220 EOS
\cite{Lattimer:1991nc}, where $t=0$ is the time of
merger~\cite{Takami2015}. Figure~\ref{fig:t_typ} uses a colorcode to show
the expansion flow timescale $t_{\rm flow} \equiv 1/\langle|\vec{\nabla}
\cdot \vec{v}|\rangle = \rho/D_t\rho$ where $\langle\ \rangle$ represents
a time average over a 2\,ms time window and where $D_t$ is the Lagrangian
time derivative in Newtonian
hydrodynamics~\cite{Rezzolla_book:2013}. This quantity is easily measured
and, for a harmonic density oscillation, it is related to
Eqs.~\eqref{bulk:tau} and \eqref{bulk:tau-min} by $t_{\rm dens} \approx
(4 \Delta n/\bar n) t_{\rm flow}$. Figure~\ref{fig:t_typ} reports $t_{\rm
  flow}$ at $2.4\,\ms$ after the merger, where the post-merger object is
in its violent and shock-dominated transient phase, (see
\cite{Takami2015} for a mechanical toy model describing this stage of the
post-merger). Inside the green contour, the rest-mass density is above
nuclear saturation. The red and grey lines are instead temperature
contours at $4\,{\rm MeV}$ and $7\,{\rm MeV}$, respectively. Overall,
Fig.~\ref{fig:t_typ} shows that there are significant regions where
Eq.~(\ref{bulk:tau-min}) is a valid estimate of the dissipation time
because the density is high and the temperature is in the range that
maximizes bulk viscosity [Eq.~(\ref{bulk:Tmax-value})]. Since in these
regions $t_{\rm flow} \sim 0.1 - 1\, {\rm ms}$ and $\Delta n/\bar n \sim
1$, we conclude that $t_{\rm dens} \approx (4 \Delta n/\bar n) t_{\rm
  flow} \sim t_{\rm flow}$, is indeed in the millisecond range.

This conclusion is reinforced by Fig.~\ref{fig:tracers}, which shows the
evolution of various local properties of representative tracer particles
in the inner region of the merger product \cite{Bovard2016}. From the top
panel, which reports the evolution of the temperature, we see that the
tracers all pass through the temperature range of large bulk viscosity
(dark and light-grey shaded areas, showing the regions of maximum and up to an order of magnitude smaller dissipation) during the first few milliseconds. The
second panel reports instead the evolution of the normalized rest-mass
density and shows that at early times (\ie $t\lesssim 5\,{\rm ms}$) there
are variations of order 100\% in the rest-mass density on a timescale of
milliseconds, confirming that $t_{\rm dens}$ is in that range. The third
panel shows the average of $t_{\rm flow}$ for the tracers, which is in
the 0.1 to 1\,ms range, as expected from Figure~\ref{fig:t_typ}. Finally,
the bottom panel of Fig.~\ref{fig:tracers} is a spectrogram averaging the
power spectral densities of the normalized rest-mass densities in the
second panel and showing how, throughout the first 20 ms, the merger
product has oscillation with a significant power at frequencies in the
kHz range.

Once again, the results shown in Figs.~\ref{fig:t_typ} and
Fig.~\ref{fig:tracers}, combined with Eq.~\eqref{bulk:tau-min}, suggest
that if direct Urca processes remain suppressed, then significant bulk
viscous dissipation may occur on timescales of order a few milliseconds, which
is fast enough to affect the flow of the nuclear material, and hence the
emitted gravitational signal. Full numerical-relativity simulations
accounting for bulk viscosity are necessary to quantify the amount of
such dissipation and its impact on the gravitational-wave signal.

\smallskip
\noindent {\em Conclusions.} Viscous dissipation or thermal transport can
play a significant role in neutron-star mergers if the relevant
dissipation time is comparable with or shorter than the survival time of
the post-merger object, which may range from tens of milliseconds for
massive stars up to seconds for lighter ones~\cite{Baiotti:2016qnr}. Our
estimates of dissipation timescales show how they depend on the
properties of the post-merger material and on the lengthscale $z_{\rm
  typ}$ over which gradients develop. We find that shear viscosity and
thermal conductivity are not likely to play a major role unless neutrino
trapping occurs, which requires $T\gtrsim 10\,{\rm MeV}$ and $z_{\rm typ}
\lesssim 0.01\,{\rm km}$. On the other hand, if direct-Urca processes
remain suppressed leaving modified-Urca processes to establish flavor
equilibrium, then bulk viscous dissipation could provide significant
damping of the high-amplitude density oscillations observed right after
merger. Data from a state-of-the-art simulation of a typical merger
confirms that the conditions for maximum bulk viscosity are present. We
conclude that viscous dissipative processes deserve more careful
investigation since they may well affect the spectral properties of the
post-merger gravitational-wave signal, especially those peaks that are
produced right after the merger and that are dissipated rapidly
\cite{Bauswein2011, Stergioulas2011b, Takami2014, Rezzolla2016,
  maione2017}. Since these peaks are routinely employed to infer the
properties of the EOS \cite{Clark2016, Bose2017}, a more realistic
treatment of the associated amplitudes is particularly important. In
addition, if viscous dissipation is active after the merger, it will also
heat the merger product, possibly stabilizing it on longer timescales via
the extra thermal pressure~\cite{Baiotti08, Sekiguchi:2011zd,
  Paschalidis:2012ff,Kaplan2013}. If future gravitational-wave
observations indicate that the actual dissipation is much smaller than
the maximum value leading to Eq.~(\ref{bulk:tau-min}), then it would be
possible to put limits on how much of the material is in phases where
direct Urca is suppressed. 

There are various directions in which our research can be further
developed. First, the effects of bulk viscosity should be consistently
included in future merger simulations. This has not been attempted before
and requires a formulation of the relativistic-hydrodynamic equations
that is hyperbolic and stable (see Chap.~6 of~\cite{Rezzolla_book:2013}
for the associated challenges). Second, the bulk viscous effects
discussed so far may be amplified by nonlinear suprathermal enhancement
\cite{Reisenegger:2003pd, Bonacic:2003th, Alford:2010gw, Alford:2011df,
  Alford:2016cee} or by the even stronger phase-conversion dissipation
\cite{Alford:2014jha}. Third, because the role played by shear viscosity
depends on the typical scale-height of the fluid flow, investigations of
the development of turbulent motion in the post-merger phase will be
essential (see \cite{Radice2017} for a first attempt).
Finally, given the vital role they play in determining the strength of
thermal transport and shear or bulk viscous dissipation, neutrino
trapping and direct-Urca processes motivate additional work to refine
constraints on the physical conditions under which these phenomena occur
in high-density matter. We plan to consider some of these topics in our
future work.

\smallskip
\noindent {\em Acknowledgements.} Support comes from: the U.S. Department
of Energy, Office of Science, Office of Nuclear Physics under Award
Number \#DE-FG02-05ER41375; ``NewCompStar'', COST Action MP1304;
LOEWE-Program in HIC for FAIR; European Union's Horizon 2020 Research and
Innovation Programme (Grant 671698) (call FETHPC-1-2014, project
ExaHyPE). It is a pleasure to thank A. Bauswein, T. Fischer,
G. McLaughlin, V. Paschalidis, S. Reddy, A. Sedrakian, S. Shapiro, and
P. Shternin for very helpful discussions. MGA and KS acknowledge the
hospitality of the Goethe University in Frankfurt, and the of Institute
for Nuclear Theory in Seattle during the program INT-16-2b ``The Phases
of Dense Matter''.

\bibliographystyle{apsrev4-1}
\bibliography{merger-dissipation,aeireferences}{}

\begin{thebibliography}{55}%
\makeatletter
\providecommand \@ifxundefined [1]{%
 \@ifx{#1\undefined}
}%
\providecommand \@ifnum [1]{%
 \ifnum #1\expandafter \@firstoftwo
 \else \expandafter \@secondoftwo
 \fi
}%
\providecommand \@ifx [1]{%
 \ifx #1\expandafter \@firstoftwo
 \else \expandafter \@secondoftwo
 \fi
}%
\providecommand \natexlab [1]{#1}%
\providecommand \enquote  [1]{``#1''}%
\providecommand \bibnamefont  [1]{#1}%
\providecommand \bibfnamefont [1]{#1}%
\providecommand \citenamefont [1]{#1}%
\providecommand \href@noop [0]{\@secondoftwo}%
\providecommand \href [0]{\begingroup \@sanitize@url \@href}%
\providecommand \@href[1]{\@@startlink{#1}\@@href}%
\providecommand \@@href[1]{\endgroup#1\@@endlink}%
\providecommand \@sanitize@url [0]{\catcode `\\12\catcode `\$12\catcode
  `\&12\catcode `\#12\catcode `\^12\catcode `\_12\catcode `\%12\relax}%
\providecommand \@@startlink[1]{}%
\providecommand \@@endlink[0]{}%
\providecommand \url  [0]{\begingroup\@sanitize@url \@url }%
\providecommand \@url [1]{\endgroup\@href {#1}{\urlprefix }}%
\providecommand \urlprefix  [0]{URL }%
\providecommand \Eprint [0]{\href }%
\providecommand \doibase [0]{http://dx.doi.org/}%
\providecommand \selectlanguage [0]{\@gobble}%
\providecommand \bibinfo  [0]{\@secondoftwo}%
\providecommand \bibfield  [0]{\@secondoftwo}%
\providecommand \translation [1]{[#1]}%
\providecommand \BibitemOpen [0]{}%
\providecommand \bibitemStop [0]{}%
\providecommand \bibitemNoStop [0]{.\EOS\space}%
\providecommand \EOS [0]{\spacefactor3000\relax}%
\providecommand \BibitemShut  [1]{\csname bibitem#1\endcsname}%
\let\auto@bib@innerbib\@empty
\bibitem [{\citenamefont {Abbott}\ \emph {et~al.}(2016)\citenamefont {Abbott}
  \emph {et~al.}}]{Abbott:2016blz}%
  \BibitemOpen
  \bibfield  {author} {\bibinfo {author} {\bibfnamefont {B.~P.}\ \bibnamefont
  {Abbott}} \emph {et~al.} (\bibinfo {collaboration} {Virgo, LIGO
  Scientific}),\ }\href {\doibase 10.1103/PhysRevLett.116.061102} {\bibfield
  {journal} {\bibinfo  {journal} {Phys. Rev. Lett.}\ }\textbf {\bibinfo
  {volume} {116}},\ \bibinfo {pages} {061102} (\bibinfo {year} {2016})},\
  \Eprint {http://arxiv.org/abs/1602.03837} {arXiv:1602.03837 [gr-qc]}
  \BibitemShut {NoStop}%
\bibitem [{\citenamefont {Baiotti}\ and\ \citenamefont
  {Rezzolla}(2016)}]{Baiotti:2016qnr}%
  \BibitemOpen
  \bibfield  {author} {\bibinfo {author} {\bibfnamefont {L.}~\bibnamefont
  {Baiotti}}\ and\ \bibinfo {author} {\bibfnamefont {L.}~\bibnamefont
  {Rezzolla}},\ }\href {\doibase 10.1088/1361-6633/aa67bb} {\bibfield
  {journal} {\bibinfo  {journal} {Reports on Progress in Physics}\ } (\bibinfo
  {year} {2016}),\ 10.1088/1361-6633/aa67bb},\ \Eprint
  {http://arxiv.org/abs/1607.03540} {arXiv:1607.03540 [gr-qc]} \BibitemShut
  {NoStop}%
\bibitem [{\citenamefont {{Paschalidis}}(2017)}]{Paschalidis2016}%
  \BibitemOpen
  \bibfield  {author} {\bibinfo {author} {\bibfnamefont {V.}~\bibnamefont
  {{Paschalidis}}},\ }\href {\doibase 10.1088/1361-6382/aa61ce} {\bibfield
  {journal} {\bibinfo  {journal} {Classical and Quantum Gravity}\ }\textbf
  {\bibinfo {volume} {34}},\ \bibinfo {eid} {084002} (\bibinfo {year}
  {2017})},\ \Eprint {http://arxiv.org/abs/1611.01519} {arXiv:1611.01519
  [astro-ph.HE]} \BibitemShut {NoStop}%
\bibitem [{\citenamefont {Duez}\ \emph {et~al.}(2004)\citenamefont {Duez},
  \citenamefont {Liu}, \citenamefont {Shapiro},\ and\ \citenamefont
  {Stephens}}]{Duez:2004nf}%
  \BibitemOpen
  \bibfield  {author} {\bibinfo {author} {\bibfnamefont {M.~D.}\ \bibnamefont
  {Duez}}, \bibinfo {author} {\bibfnamefont {Y.~T.}\ \bibnamefont {Liu}},
  \bibinfo {author} {\bibfnamefont {S.~L.}\ \bibnamefont {Shapiro}}, \ and\
  \bibinfo {author} {\bibfnamefont {B.~C.}\ \bibnamefont {Stephens}},\ }\href
  {\doibase 10.1103/PhysRevD.69.104030} {\bibfield  {journal} {\bibinfo
  {journal} {Phys. Rev.}\ }\textbf {\bibinfo {volume} {D69}},\ \bibinfo {pages}
  {104030} (\bibinfo {year} {2004})},\ \Eprint
  {http://arxiv.org/abs/astro-ph/0402502} {arXiv:astro-ph/0402502 [astro-ph]}
  \BibitemShut {NoStop}%
\bibitem [{\citenamefont {{Shibata}}\ and\ \citenamefont
  {{Kiuchi}}(2017)}]{Shibata:2017b}%
  \BibitemOpen
  \bibfield  {author} {\bibinfo {author} {\bibfnamefont {M.}~\bibnamefont
  {{Shibata}}}\ and\ \bibinfo {author} {\bibfnamefont {K.}~\bibnamefont
  {{Kiuchi}}},\ }\href@noop {} {\bibfield  {journal} {\bibinfo  {journal}
  {arXiv:1705.06142}\ } (\bibinfo {year} {2017})},\ \Eprint
  {http://arxiv.org/abs/1705.06142} {arXiv:1705.06142 [astro-ph.HE]}
  \BibitemShut {NoStop}%
\bibitem [{\citenamefont {{Bildsten}}\ and\ \citenamefont
  {{Cutler}}(1992)}]{Bildsten92}%
  \BibitemOpen
  \bibfield  {author} {\bibinfo {author} {\bibfnamefont {L.}~\bibnamefont
  {{Bildsten}}}\ and\ \bibinfo {author} {\bibfnamefont {C.}~\bibnamefont
  {{Cutler}}},\ }\href {\doibase 10.1086/171983} {\bibfield  {journal}
  {\bibinfo  {journal} {Astrophys. J.}\ }\textbf {\bibinfo {volume} {400}},\
  \bibinfo {pages} {175} (\bibinfo {year} {1992})}\BibitemShut {NoStop}%
\bibitem [{\citenamefont {{Shibata}}\ and\ \citenamefont
  {{Ury{\=u}}}(2000)}]{Shibata99d}%
  \BibitemOpen
  \bibfield  {author} {\bibinfo {author} {\bibfnamefont {M.}~\bibnamefont
  {{Shibata}}}\ and\ \bibinfo {author} {\bibfnamefont {K.}~\bibnamefont
  {{Ury{\=u}}}},\ }\href {\doibase 10.1103/PhysRevD.61.064001} {\bibfield
  {journal} {\bibinfo  {journal} {Phys. Rev. D}\ }\textbf {\bibinfo {volume}
  {61}},\ \bibinfo {eid} {064001} (\bibinfo {year} {2000})},\ \Eprint
  {http://arxiv.org/abs/gr-qc/9911058} {gr-qc/9911058} \BibitemShut {NoStop}%
\bibitem [{\citenamefont {{Baiotti}}\ \emph {et~al.}(2008)\citenamefont
  {{Baiotti}}, \citenamefont {{Giacomazzo}},\ and\ \citenamefont
  {{Rezzolla}}}]{Baiotti08}%
  \BibitemOpen
  \bibfield  {author} {\bibinfo {author} {\bibfnamefont {L.}~\bibnamefont
  {{Baiotti}}}, \bibinfo {author} {\bibfnamefont {B.}~\bibnamefont
  {{Giacomazzo}}}, \ and\ \bibinfo {author} {\bibfnamefont {L.}~\bibnamefont
  {{Rezzolla}}},\ }\href {\doibase 10.1103/PhysRevD.78.084033} {\bibfield
  {journal} {\bibinfo  {journal} {Phys. Rev. D}\ }\textbf {\bibinfo {volume}
  {78}},\ \bibinfo {pages} {084033} (\bibinfo {year} {2008})},\ \Eprint
  {http://arxiv.org/abs/0804.0594} {arXiv:0804.0594 [gr-qc]} \BibitemShut
  {NoStop}%
\bibitem [{\citenamefont {{Anderson}}\ \emph {et~al.}(2008)\citenamefont
  {{Anderson}}, \citenamefont {{Hirschmann}}, \citenamefont {{Lehner}},
  \citenamefont {{Liebling}}, \citenamefont {{Motl}}, \citenamefont
  {{Neilsen}}, \citenamefont {{Palenzuela}},\ and\ \citenamefont
  {{Tohline}}}]{Anderson2007}%
  \BibitemOpen
  \bibfield  {author} {\bibinfo {author} {\bibfnamefont {M.}~\bibnamefont
  {{Anderson}}}, \bibinfo {author} {\bibfnamefont {E.~W.}\ \bibnamefont
  {{Hirschmann}}}, \bibinfo {author} {\bibfnamefont {L.}~\bibnamefont
  {{Lehner}}}, \bibinfo {author} {\bibfnamefont {S.~L.}\ \bibnamefont
  {{Liebling}}}, \bibinfo {author} {\bibfnamefont {P.~M.}\ \bibnamefont
  {{Motl}}}, \bibinfo {author} {\bibfnamefont {D.}~\bibnamefont {{Neilsen}}},
  \bibinfo {author} {\bibfnamefont {C.}~\bibnamefont {{Palenzuela}}}, \ and\
  \bibinfo {author} {\bibfnamefont {J.~E.}\ \bibnamefont {{Tohline}}},\ }\href
  {\doibase 10.1103/PhysRevD.77.024006} {\bibfield  {journal} {\bibinfo
  {journal} {Phys. Rev. D}\ }\textbf {\bibinfo {volume} {77}},\ \bibinfo {eid}
  {024006} (\bibinfo {year} {2008})},\ \Eprint {http://arxiv.org/abs/0708.2720}
  {arXiv:0708.2720 [gr-qc]} \BibitemShut {NoStop}%
\bibitem [{\citenamefont {Liu}\ \emph {et~al.}(2008)\citenamefont {Liu},
  \citenamefont {Shapiro}, \citenamefont {Etienne},\ and\ \citenamefont
  {Taniguchi}}]{Liu:2008xy}%
  \BibitemOpen
  \bibfield  {author} {\bibinfo {author} {\bibfnamefont {Y.~T.}\ \bibnamefont
  {Liu}}, \bibinfo {author} {\bibfnamefont {S.~L.}\ \bibnamefont {Shapiro}},
  \bibinfo {author} {\bibfnamefont {Z.~B.}\ \bibnamefont {Etienne}}, \ and\
  \bibinfo {author} {\bibfnamefont {K.}~\bibnamefont {Taniguchi}},\ }\href
  {\doibase 10.1103/PhysRevD.78.024012} {\bibfield  {journal} {\bibinfo
  {journal} {Phys. Rev. D}\ }\textbf {\bibinfo {volume} {78}},\ \bibinfo
  {pages} {024012} (\bibinfo {year} {2008})},\ \Eprint
  {http://arxiv.org/abs/0803.4193} {arXiv:0803.4193 [astro-ph]} \BibitemShut
  {NoStop}%
\bibitem [{\citenamefont {{Bernuzzi}}\ \emph {et~al.}(2012)\citenamefont
  {{Bernuzzi}}, \citenamefont {{Thierfelder}},\ and\ \citenamefont
  {{Br{\"u}gmann}}}]{Bernuzzi2011}%
  \BibitemOpen
  \bibfield  {author} {\bibinfo {author} {\bibfnamefont {S.}~\bibnamefont
  {{Bernuzzi}}}, \bibinfo {author} {\bibfnamefont {M.}~\bibnamefont
  {{Thierfelder}}}, \ and\ \bibinfo {author} {\bibfnamefont {B.}~\bibnamefont
  {{Br{\"u}gmann}}},\ }\href {\doibase 10.1103/PhysRevD.85.104030} {\bibfield
  {journal} {\bibinfo  {journal} {Phys. Rev. D}\ }\textbf {\bibinfo {volume}
  {85}},\ \bibinfo {eid} {104030} (\bibinfo {year} {2012})},\ \Eprint
  {http://arxiv.org/abs/1109.3611} {arXiv:1109.3611 [gr-qc]} \BibitemShut
  {NoStop}%
\bibitem [{\citenamefont {{Bauswein}}\ and\ \citenamefont
  {{Janka}}(2012)}]{Bauswein2011}%
  \BibitemOpen
  \bibfield  {author} {\bibinfo {author} {\bibfnamefont {A.}~\bibnamefont
  {{Bauswein}}}\ and\ \bibinfo {author} {\bibfnamefont {H.-T.}\ \bibnamefont
  {{Janka}}},\ }\href {\doibase 10.1103/PhysRevLett.108.011101} {\bibfield
  {journal} {\bibinfo  {journal} {Phys. Rev. Lett.}\ }\textbf {\bibinfo
  {volume} {108}},\ \bibinfo {eid} {011101} (\bibinfo {year} {2012})},\ \Eprint
  {http://arxiv.org/abs/1106.1616} {arXiv:1106.1616 [astro-ph.SR]} \BibitemShut
  {NoStop}%
\bibitem [{\citenamefont {{Stergioulas}}\ \emph {et~al.}(2011)\citenamefont
  {{Stergioulas}}, \citenamefont {{Bauswein}}, \citenamefont {{Zagkouris}},\
  and\ \citenamefont {{Janka}}}]{Stergioulas2011b}%
  \BibitemOpen
  \bibfield  {author} {\bibinfo {author} {\bibfnamefont {N.}~\bibnamefont
  {{Stergioulas}}}, \bibinfo {author} {\bibfnamefont {A.}~\bibnamefont
  {{Bauswein}}}, \bibinfo {author} {\bibfnamefont {K.}~\bibnamefont
  {{Zagkouris}}}, \ and\ \bibinfo {author} {\bibfnamefont {H.-T.}\ \bibnamefont
  {{Janka}}},\ }\href {\doibase 10.1111/j.1365-2966.2011.19493.x} {\bibfield
  {journal} {\bibinfo  {journal} {Mon. Not. R. Astron. Soc.}\ }\textbf
  {\bibinfo {volume} {418}},\ \bibinfo {pages} {427} (\bibinfo {year}
  {2011})},\ \Eprint {http://arxiv.org/abs/1105.0368} {arXiv:1105.0368 [gr-qc]}
  \BibitemShut {NoStop}%
\bibitem [{\citenamefont {{Takami}}\ \emph {et~al.}(2014)\citenamefont
  {{Takami}}, \citenamefont {{Rezzolla}},\ and\ \citenamefont
  {{Baiotti}}}]{Takami2014}%
  \BibitemOpen
  \bibfield  {author} {\bibinfo {author} {\bibfnamefont {K.}~\bibnamefont
  {{Takami}}}, \bibinfo {author} {\bibfnamefont {L.}~\bibnamefont
  {{Rezzolla}}}, \ and\ \bibinfo {author} {\bibfnamefont {L.}~\bibnamefont
  {{Baiotti}}},\ }\href {\doibase 10.1103/PhysRevLett.113.091104} {\bibfield
  {journal} {\bibinfo  {journal} {Phys. Rev. Lett.}\ }\textbf {\bibinfo
  {volume} {113}},\ \bibinfo {eid} {091104} (\bibinfo {year} {2014})},\ \Eprint
  {http://arxiv.org/abs/1403.5672} {arXiv:1403.5672 [gr-qc]} \BibitemShut
  {NoStop}%
\bibitem [{\citenamefont {{Bernuzzi}}\ \emph {et~al.}(2015)\citenamefont
  {{Bernuzzi}}, \citenamefont {{Dietrich}},\ and\ \citenamefont
  {{Nagar}}}]{Bernuzzi2015a}%
  \BibitemOpen
  \bibfield  {author} {\bibinfo {author} {\bibfnamefont {S.}~\bibnamefont
  {{Bernuzzi}}}, \bibinfo {author} {\bibfnamefont {T.}~\bibnamefont
  {{Dietrich}}}, \ and\ \bibinfo {author} {\bibfnamefont {A.}~\bibnamefont
  {{Nagar}}},\ }\href {\doibase 10.1103/PhysRevLett.115.091101} {\bibfield
  {journal} {\bibinfo  {journal} {Phys. Rev. Lett.}\ }\textbf {\bibinfo
  {volume} {115}},\ \bibinfo {eid} {091101} (\bibinfo {year} {2015})},\ \Eprint
  {http://arxiv.org/abs/1504.01764} {arXiv:1504.01764 [gr-qc]} \BibitemShut
  {NoStop}%
\bibitem [{\citenamefont {{Rezzolla}}\ and\ \citenamefont
  {{Takami}}(2016)}]{Rezzolla2016}%
  \BibitemOpen
  \bibfield  {author} {\bibinfo {author} {\bibfnamefont {L.}~\bibnamefont
  {{Rezzolla}}}\ and\ \bibinfo {author} {\bibfnamefont {K.}~\bibnamefont
  {{Takami}}},\ }\href {\doibase 10.1103/PhysRevD.93.124051} {\bibfield
  {journal} {\bibinfo  {journal} {Phys. Rev. D}\ }\textbf {\bibinfo {volume}
  {93}},\ \bibinfo {eid} {124051} (\bibinfo {year} {2016})},\ \Eprint
  {http://arxiv.org/abs/1604.00246} {arXiv:1604.00246 [gr-qc]} \BibitemShut
  {NoStop}%
\bibitem [{\citenamefont {{Levenfish}}\ and\ \citenamefont
  {{Yakovlev}}(1994)}]{Levenfish:1994}%
  \BibitemOpen
  \bibfield  {author} {\bibinfo {author} {\bibfnamefont {K.~P.}\ \bibnamefont
  {{Levenfish}}}\ and\ \bibinfo {author} {\bibfnamefont {D.~G.}\ \bibnamefont
  {{Yakovlev}}},\ }\href@noop {} {\bibfield  {journal} {\bibinfo  {journal}
  {Astronomy Reports}\ }\textbf {\bibinfo {volume} {38}},\ \bibinfo {pages}
  {247} (\bibinfo {year} {1994})}\BibitemShut {NoStop}%
\bibitem [{\citenamefont {Reddy}\ \emph {et~al.}(1998)\citenamefont {Reddy},
  \citenamefont {Prakash},\ and\ \citenamefont {Lattimer}}]{Reddy:1997yr}%
  \BibitemOpen
  \bibfield  {author} {\bibinfo {author} {\bibfnamefont {S.}~\bibnamefont
  {Reddy}}, \bibinfo {author} {\bibfnamefont {M.}~\bibnamefont {Prakash}}, \
  and\ \bibinfo {author} {\bibfnamefont {J.~M.}\ \bibnamefont {Lattimer}},\
  }\href {\doibase 10.1103/PhysRevD.58.013009} {\bibfield  {journal} {\bibinfo
  {journal} {Phys. Rev.}\ }\textbf {\bibinfo {volume} {D58}},\ \bibinfo {pages}
  {013009} (\bibinfo {year} {1998})},\ \Eprint
  {http://arxiv.org/abs/astro-ph/9710115} {arXiv:astro-ph/9710115 [astro-ph]}
  \BibitemShut {NoStop}%
\bibitem [{\citenamefont {Shternin}\ and\ \citenamefont
  {Yakovlev}(2007)}]{Shternin:2007ee}%
  \BibitemOpen
  \bibfield  {author} {\bibinfo {author} {\bibfnamefont {P.}~\bibnamefont
  {Shternin}}\ and\ \bibinfo {author} {\bibfnamefont {D.}~\bibnamefont
  {Yakovlev}},\ }\href {\doibase 10.1103/PhysRevD.75.103004} {\bibfield
  {journal} {\bibinfo  {journal} {Phys.Rev.}\ }\textbf {\bibinfo {volume}
  {D75}},\ \bibinfo {pages} {103004} (\bibinfo {year} {2007})},\ \Eprint
  {http://arxiv.org/abs/0705.1963} {arXiv:0705.1963 [astro-ph]} \BibitemShut
  {NoStop}%
\bibitem [{\citenamefont {Roberts}\ and\ \citenamefont
  {Reddy}(2017)}]{Roberts:2016mwj}%
  \BibitemOpen
  \bibfield  {author} {\bibinfo {author} {\bibfnamefont {L.~F.}\ \bibnamefont
  {Roberts}}\ and\ \bibinfo {author} {\bibfnamefont {S.}~\bibnamefont
  {Reddy}},\ }\href {\doibase 10.1103/PhysRevC.95.045807} {\bibfield  {journal}
  {\bibinfo  {journal} {Phys. Rev.}\ }\textbf {\bibinfo {volume} {C95}},\
  \bibinfo {pages} {045807} (\bibinfo {year} {2017})},\ \Eprint
  {http://arxiv.org/abs/1612.02764} {arXiv:1612.02764 [astro-ph.HE]}
  \BibitemShut {NoStop}%
\bibitem [{\citenamefont {Goodwin}\ and\ \citenamefont
  {Pethick}(1982)}]{Goodwin:1982hy}%
  \BibitemOpen
  \bibfield  {author} {\bibinfo {author} {\bibfnamefont {B.~T.}\ \bibnamefont
  {Goodwin}}\ and\ \bibinfo {author} {\bibfnamefont {C.~J.}\ \bibnamefont
  {Pethick}},\ }\href {\doibase 10.1086/159684} {\bibfield  {journal} {\bibinfo
   {journal} {Astrophys. J.}\ }\textbf {\bibinfo {volume} {253}},\ \bibinfo
  {pages} {816} (\bibinfo {year} {1982})}\BibitemShut {NoStop}%
\bibitem [{\citenamefont {Shternin}\ and\ \citenamefont
  {Yakovlev}(2008)}]{Shternin:2008es}%
  \BibitemOpen
  \bibfield  {author} {\bibinfo {author} {\bibfnamefont {P.~S.}\ \bibnamefont
  {Shternin}}\ and\ \bibinfo {author} {\bibfnamefont {D.~G.}\ \bibnamefont
  {Yakovlev}},\ }\href {\doibase 10.1103/PhysRevD.78.063006} {\bibfield
  {journal} {\bibinfo  {journal} {Phys. Rev.}\ }\textbf {\bibinfo {volume}
  {D78}},\ \bibinfo {pages} {063006} (\bibinfo {year} {2008})},\ \Eprint
  {http://arxiv.org/abs/0808.2018} {arXiv:0808.2018 [astro-ph]} \BibitemShut
  {NoStop}%
\bibitem [{\citenamefont {Manuel}\ and\ \citenamefont
  {Tolos}(2013)}]{Manuel:2012rd}%
  \BibitemOpen
  \bibfield  {author} {\bibinfo {author} {\bibfnamefont {C.}~\bibnamefont
  {Manuel}}\ and\ \bibinfo {author} {\bibfnamefont {L.}~\bibnamefont {Tolos}},\
  }\href {\doibase 10.1103/PhysRevD.88.043001} {\bibfield  {journal} {\bibinfo
  {journal} {Phys. Rev.}\ }\textbf {\bibinfo {volume} {D88}},\ \bibinfo {pages}
  {043001} (\bibinfo {year} {2013})},\ \Eprint {http://arxiv.org/abs/1212.2075}
  {arXiv:1212.2075 [astro-ph.SR]} \BibitemShut {NoStop}%
\bibitem [{\citenamefont {Kiuchi}\ \emph {et~al.}(2015)\citenamefont {Kiuchi},
  \citenamefont {Cerd\'a-Dur\'an}, \citenamefont {Kyutoku}, \citenamefont
  {Sekiguchi},\ and\ \citenamefont {Shibata}}]{Kiuchi:2015sga}%
  \BibitemOpen
  \bibfield  {author} {\bibinfo {author} {\bibfnamefont {K.}~\bibnamefont
  {Kiuchi}}, \bibinfo {author} {\bibfnamefont {P.}~\bibnamefont
  {Cerd\'a-Dur\'an}}, \bibinfo {author} {\bibfnamefont {K.}~\bibnamefont
  {Kyutoku}}, \bibinfo {author} {\bibfnamefont {Y.}~\bibnamefont {Sekiguchi}},
  \ and\ \bibinfo {author} {\bibfnamefont {M.}~\bibnamefont {Shibata}},\ }\href
  {\doibase 10.1103/PhysRevD.92.124034} {\bibfield  {journal} {\bibinfo
  {journal} {Phys. Rev.}\ }\textbf {\bibinfo {volume} {D92}},\ \bibinfo {pages}
  {124034} (\bibinfo {year} {2015})},\ \Eprint
  {http://arxiv.org/abs/1509.09205} {arXiv:1509.09205 [astro-ph.HE]}
  \BibitemShut {NoStop}%
\bibitem [{\citenamefont {{East}}\ \emph {et~al.}(2016)\citenamefont {{East}},
  \citenamefont {{Paschalidis}}, \citenamefont {{Pretorius}},\ and\
  \citenamefont {{Shapiro}}}]{East2016}%
  \BibitemOpen
  \bibfield  {author} {\bibinfo {author} {\bibfnamefont {W.~E.}\ \bibnamefont
  {{East}}}, \bibinfo {author} {\bibfnamefont {V.}~\bibnamefont
  {{Paschalidis}}}, \bibinfo {author} {\bibfnamefont {F.}~\bibnamefont
  {{Pretorius}}}, \ and\ \bibinfo {author} {\bibfnamefont {S.~L.}\ \bibnamefont
  {{Shapiro}}},\ }\href {\doibase 10.1103/PhysRevD.93.024011} {\bibfield
  {journal} {\bibinfo  {journal} {Phys. Rev. D}\ }\textbf {\bibinfo {volume}
  {93}},\ \bibinfo {eid} {024011} (\bibinfo {year} {2016})},\ \Eprint
  {http://arxiv.org/abs/1511.01093} {arXiv:1511.01093 [astro-ph.HE]}
  \BibitemShut {NoStop}%
\bibitem [{\citenamefont {{Radice}}\ \emph {et~al.}(2016)\citenamefont
  {{Radice}}, \citenamefont {{Bernuzzi}},\ and\ \citenamefont
  {{Ott}}}]{Radice2016a}%
  \BibitemOpen
  \bibfield  {author} {\bibinfo {author} {\bibfnamefont {D.}~\bibnamefont
  {{Radice}}}, \bibinfo {author} {\bibfnamefont {S.}~\bibnamefont
  {{Bernuzzi}}}, \ and\ \bibinfo {author} {\bibfnamefont {C.~D.}\ \bibnamefont
  {{Ott}}},\ }\href {\doibase 10.1103/PhysRevD.94.064011} {\bibfield  {journal}
  {\bibinfo  {journal} {Phys. Rev. D}\ }\textbf {\bibinfo {volume} {94}},\
  \bibinfo {eid} {064011} (\bibinfo {year} {2016})},\ \Eprint
  {http://arxiv.org/abs/1603.05726} {arXiv:1603.05726 [gr-qc]} \BibitemShut
  {NoStop}%
\bibitem [{\citenamefont {{Lehner}}\ \emph {et~al.}(2016)\citenamefont
  {{Lehner}}, \citenamefont {{Liebling}}, \citenamefont {{Palenzuela}},\ and\
  \citenamefont {{Motl}}}]{Lehner2016a}%
  \BibitemOpen
  \bibfield  {author} {\bibinfo {author} {\bibfnamefont {L.}~\bibnamefont
  {{Lehner}}}, \bibinfo {author} {\bibfnamefont {S.~L.}\ \bibnamefont
  {{Liebling}}}, \bibinfo {author} {\bibfnamefont {C.}~\bibnamefont
  {{Palenzuela}}}, \ and\ \bibinfo {author} {\bibfnamefont {P.~M.}\
  \bibnamefont {{Motl}}},\ }\href {\doibase 10.1103/PhysRevD.94.043003}
  {\bibfield  {journal} {\bibinfo  {journal} {Phys. Rev. D}\ }\textbf {\bibinfo
  {volume} {94}},\ \bibinfo {eid} {043003} (\bibinfo {year} {2016})},\ \Eprint
  {http://arxiv.org/abs/1605.02369} {arXiv:1605.02369 [gr-qc]} \BibitemShut
  {NoStop}%
\bibitem [{\citenamefont {Alford}\ \emph {et~al.}(2010)\citenamefont {Alford},
  \citenamefont {Mahmoodifar},\ and\ \citenamefont
  {Schwenzer}}]{Alford:2010gw}%
  \BibitemOpen
  \bibfield  {author} {\bibinfo {author} {\bibfnamefont {M.~G.}\ \bibnamefont
  {Alford}}, \bibinfo {author} {\bibfnamefont {S.}~\bibnamefont {Mahmoodifar}},
  \ and\ \bibinfo {author} {\bibfnamefont {K.}~\bibnamefont {Schwenzer}},\
  }\href {\doibase 10.1088/0954-3899/37/12/125202} {\bibfield  {journal}
  {\bibinfo  {journal} {J. Phys.}\ }\textbf {\bibinfo {volume} {G37}},\
  \bibinfo {pages} {125202} (\bibinfo {year} {2010})},\ \Eprint
  {http://arxiv.org/abs/1005.3769} {arXiv:1005.3769 [nucl-th]} \BibitemShut
  {NoStop}%
\bibitem [{\citenamefont {Demorest}\ \emph {et~al.}(2010)\citenamefont
  {Demorest}, \citenamefont {Pennucci}, \citenamefont {Ransom}, \citenamefont
  {Roberts},\ and\ \citenamefont {Hessels}}]{Demorest:2010bx}%
  \BibitemOpen
  \bibfield  {author} {\bibinfo {author} {\bibfnamefont {P.}~\bibnamefont
  {Demorest}}, \bibinfo {author} {\bibfnamefont {T.}~\bibnamefont {Pennucci}},
  \bibinfo {author} {\bibfnamefont {S.}~\bibnamefont {Ransom}}, \bibinfo
  {author} {\bibfnamefont {M.}~\bibnamefont {Roberts}}, \ and\ \bibinfo
  {author} {\bibfnamefont {J.}~\bibnamefont {Hessels}},\ }\href@noop {}
  {\bibfield  {journal} {\bibinfo  {journal} {Nature}\ }\textbf {\bibinfo
  {volume} {467}},\ \bibinfo {pages} {1081} (\bibinfo {year} {2010})},\ \Eprint
  {http://arxiv.org/abs/1010.5788} {arXiv:1010.5788 [astro-ph.HE]} \BibitemShut
  {NoStop}%
\bibitem [{\citenamefont {{Antoniadis}}\ \emph {et~al.}(2013)\citenamefont
  {{Antoniadis}}, \citenamefont {{Freire}}, \citenamefont {{Wex}},
  \citenamefont {{Tauris}}, \citenamefont {{Lynch}}, \citenamefont {{van
  Kerkwijk}}, \citenamefont {{Kramer}}, \citenamefont {{Bassa}}, \citenamefont
  {{Dhillon}}, \citenamefont {{Driebe}}, \citenamefont {{Hessels}},
  \citenamefont {{Kaspi}}, \citenamefont {{Kondratiev}}, \citenamefont
  {{Langer}}, \citenamefont {{Marsh}}, \citenamefont {{McLaughlin}},
  \citenamefont {{Pennucci}}, \citenamefont {{Ransom}}, \citenamefont
  {{Stairs}}, \citenamefont {{van Leeuwen}}, \citenamefont {{Verbiest}},\ and\
  \citenamefont {{Whelan}}}]{Antoniadis2013}%
  \BibitemOpen
  \bibfield  {author} {\bibinfo {author} {\bibfnamefont {J.}~\bibnamefont
  {{Antoniadis}}}, \bibinfo {author} {\bibfnamefont {P.~C.~C.}\ \bibnamefont
  {{Freire}}}, \bibinfo {author} {\bibfnamefont {N.}~\bibnamefont {{Wex}}},
  \bibinfo {author} {\bibfnamefont {T.~M.}\ \bibnamefont {{Tauris}}}, \bibinfo
  {author} {\bibfnamefont {R.~S.}\ \bibnamefont {{Lynch}}}, \bibinfo {author}
  {\bibfnamefont {M.~H.}\ \bibnamefont {{van Kerkwijk}}}, \bibinfo {author}
  {\bibfnamefont {M.}~\bibnamefont {{Kramer}}}, \bibinfo {author}
  {\bibfnamefont {C.}~\bibnamefont {{Bassa}}}, \bibinfo {author} {\bibfnamefont
  {V.~S.}\ \bibnamefont {{Dhillon}}}, \bibinfo {author} {\bibfnamefont
  {T.}~\bibnamefont {{Driebe}}}, \bibinfo {author} {\bibfnamefont {J.~W.~T.}\
  \bibnamefont {{Hessels}}}, \bibinfo {author} {\bibfnamefont {V.~M.}\
  \bibnamefont {{Kaspi}}}, \bibinfo {author} {\bibfnamefont {V.~I.}\
  \bibnamefont {{Kondratiev}}}, \bibinfo {author} {\bibfnamefont
  {N.}~\bibnamefont {{Langer}}}, \bibinfo {author} {\bibfnamefont {T.~R.}\
  \bibnamefont {{Marsh}}}, \bibinfo {author} {\bibfnamefont {M.~A.}\
  \bibnamefont {{McLaughlin}}}, \bibinfo {author} {\bibfnamefont {T.~T.}\
  \bibnamefont {{Pennucci}}}, \bibinfo {author} {\bibfnamefont {S.~M.}\
  \bibnamefont {{Ransom}}}, \bibinfo {author} {\bibfnamefont {I.~H.}\
  \bibnamefont {{Stairs}}}, \bibinfo {author} {\bibfnamefont {J.}~\bibnamefont
  {{van Leeuwen}}}, \bibinfo {author} {\bibfnamefont {J.~P.~W.}\ \bibnamefont
  {{Verbiest}}}, \ and\ \bibinfo {author} {\bibfnamefont {D.~G.}\ \bibnamefont
  {{Whelan}}},\ }\href {\doibase 10.1126/science.1233232} {\bibfield  {journal}
  {\bibinfo  {journal} {Science}\ }\textbf {\bibinfo {volume} {340}},\ \bibinfo
  {pages} {448} (\bibinfo {year} {2013})},\ \Eprint
  {http://arxiv.org/abs/1304.6875} {arXiv:1304.6875 [astro-ph.HE]} \BibitemShut
  {NoStop}%
\bibitem [{\citenamefont {Akmal}\ \emph {et~al.}(1998)\citenamefont {Akmal},
  \citenamefont {Pandharipande},\ and\ \citenamefont
  {Ravenhall}}]{Akmal:1998cf}%
  \BibitemOpen
  \bibfield  {author} {\bibinfo {author} {\bibfnamefont {A.}~\bibnamefont
  {Akmal}}, \bibinfo {author} {\bibfnamefont {V.~R.}\ \bibnamefont
  {Pandharipande}}, \ and\ \bibinfo {author} {\bibfnamefont {D.~G.}\
  \bibnamefont {Ravenhall}},\ }\href {\doibase 10.1103/PhysRevC.58.1804}
  {\bibfield  {journal} {\bibinfo  {journal} {Phys. Rev.}\ }\textbf {\bibinfo
  {volume} {C58}},\ \bibinfo {pages} {1804} (\bibinfo {year} {1998})},\ \Eprint
  {http://arxiv.org/abs/nucl-th/9804027} {arXiv:nucl-th/9804027} \BibitemShut
  {NoStop}%
\bibitem [{\citenamefont {Hempel}\ and\ \citenamefont
  {Schaffner-Bielich}(2010)}]{Hempel:2009mc}%
  \BibitemOpen
  \bibfield  {author} {\bibinfo {author} {\bibfnamefont {M.}~\bibnamefont
  {Hempel}}\ and\ \bibinfo {author} {\bibfnamefont {J.}~\bibnamefont
  {Schaffner-Bielich}},\ }\href {\doibase 10.1016/j.nuclphysa.2010.02.010}
  {\bibfield  {journal} {\bibinfo  {journal} {Nucl. Phys.}\ }\textbf {\bibinfo
  {volume} {A837}},\ \bibinfo {pages} {210} (\bibinfo {year} {2010})},\ \Eprint
  {http://arxiv.org/abs/0911.4073} {arXiv:0911.4073 [nucl-th]} \BibitemShut
  {NoStop}%
\bibitem [{\citenamefont {Lattimer}\ and\ \citenamefont
  {Swesty}(1991)}]{Lattimer:1991nc}%
  \BibitemOpen
  \bibfield  {author} {\bibinfo {author} {\bibfnamefont {J.~M.}\ \bibnamefont
  {Lattimer}}\ and\ \bibinfo {author} {\bibfnamefont {F.~D.}\ \bibnamefont
  {Swesty}},\ }\href {\doibase 10.1016/0375-9474(91)90452-C} {\bibfield
  {journal} {\bibinfo  {journal} {Nucl. Phys.}\ }\textbf {\bibinfo {volume}
  {A535}},\ \bibinfo {pages} {331} (\bibinfo {year} {1991})}\BibitemShut
  {NoStop}%
\bibitem [{\citenamefont {Steiner}\ \emph {et~al.}(2013)\citenamefont
  {Steiner}, \citenamefont {Hempel},\ and\ \citenamefont
  {Fischer}}]{Steiner:2012rk}%
  \BibitemOpen
  \bibfield  {author} {\bibinfo {author} {\bibfnamefont {A.~W.}\ \bibnamefont
  {Steiner}}, \bibinfo {author} {\bibfnamefont {M.}~\bibnamefont {Hempel}}, \
  and\ \bibinfo {author} {\bibfnamefont {T.}~\bibnamefont {Fischer}},\ }\href
  {\doibase 10.1088/0004-637X/774/1/17} {\bibfield  {journal} {\bibinfo
  {journal} {Astrophys. J.}\ }\textbf {\bibinfo {volume} {774}},\ \bibinfo
  {pages} {17} (\bibinfo {year} {2013})},\ \Eprint
  {http://arxiv.org/abs/1207.2184} {arXiv:1207.2184 [astro-ph.SR]} \BibitemShut
  {NoStop}%
\bibitem [{\citenamefont {Fischer}\ \emph {et~al.}(2014)\citenamefont
  {Fischer}, \citenamefont {Hempel}, \citenamefont {Sagert}, \citenamefont
  {Suwa},\ and\ \citenamefont {Schaffner-Bielich}}]{Fischer:2013eka}%
  \BibitemOpen
  \bibfield  {author} {\bibinfo {author} {\bibfnamefont {T.}~\bibnamefont
  {Fischer}}, \bibinfo {author} {\bibfnamefont {M.}~\bibnamefont {Hempel}},
  \bibinfo {author} {\bibfnamefont {I.}~\bibnamefont {Sagert}}, \bibinfo
  {author} {\bibfnamefont {Y.}~\bibnamefont {Suwa}}, \ and\ \bibinfo {author}
  {\bibfnamefont {J.}~\bibnamefont {Schaffner-Bielich}},\ }\href {\doibase
  10.1140/epja/i2014-14046-5} {\bibfield  {journal} {\bibinfo  {journal} {Eur.
  Phys. J.}\ }\textbf {\bibinfo {volume} {A50}},\ \bibinfo {pages} {46}
  (\bibinfo {year} {2014})},\ \Eprint {http://arxiv.org/abs/1307.6190}
  {arXiv:1307.6190 [astro-ph.HE]} \BibitemShut {NoStop}%
\bibitem [{\citenamefont {{Dietrich}}\ and\ \citenamefont
  {{Ujevic}}(2016)}]{Dietrich2016}%
  \BibitemOpen
  \bibfield  {author} {\bibinfo {author} {\bibfnamefont {T.}~\bibnamefont
  {{Dietrich}}}\ and\ \bibinfo {author} {\bibfnamefont {M.}~\bibnamefont
  {{Ujevic}}},\ }\href@noop {} {\bibfield  {journal} {\bibinfo  {journal}
  {ArXiv e-prints}\ } (\bibinfo {year} {2016})},\ \Eprint
  {http://arxiv.org/abs/1612.03665} {arXiv:1612.03665 [gr-qc]} \BibitemShut
  {NoStop}%
\bibitem [{\citenamefont {{Hanauske}}\ \emph {et~al.}(2016)\citenamefont
  {{Hanauske}}, \citenamefont {{Takami}}, \citenamefont {{Bovard}},
  \citenamefont {{Rezzolla}}, \citenamefont {{Font}}, \citenamefont
  {{Galeazzi}},\ and\ \citenamefont {{St{\"o}cker}}}]{Hanauske2016}%
  \BibitemOpen
  \bibfield  {author} {\bibinfo {author} {\bibfnamefont {M.}~\bibnamefont
  {{Hanauske}}}, \bibinfo {author} {\bibfnamefont {K.}~\bibnamefont
  {{Takami}}}, \bibinfo {author} {\bibfnamefont {L.}~\bibnamefont {{Bovard}}},
  \bibinfo {author} {\bibfnamefont {L.}~\bibnamefont {{Rezzolla}}}, \bibinfo
  {author} {\bibfnamefont {J.~A.}\ \bibnamefont {{Font}}}, \bibinfo {author}
  {\bibfnamefont {F.}~\bibnamefont {{Galeazzi}}}, \ and\ \bibinfo {author}
  {\bibfnamefont {H.}~\bibnamefont {{St{\"o}cker}}},\ }\href@noop {} {\bibfield
   {journal} {\bibinfo  {journal} {ArXiv e-prints}\ } (\bibinfo {year}
  {2016})},\ \Eprint {http://arxiv.org/abs/1611.07152} {arXiv:1611.07152
  [gr-qc]} \BibitemShut {NoStop}%
\bibitem [{\citenamefont {Klahn}\ \emph {et~al.}(2006)\citenamefont {Klahn}
  \emph {et~al.}}]{Klahn:2006ir}%
  \BibitemOpen
  \bibfield  {author} {\bibinfo {author} {\bibfnamefont {T.}~\bibnamefont
  {Klahn}} \emph {et~al.},\ }\href {\doibase 10.1103/PhysRevC.74.035802}
  {\bibfield  {journal} {\bibinfo  {journal} {Phys. Rev.}\ }\textbf {\bibinfo
  {volume} {C74}},\ \bibinfo {pages} {035802} (\bibinfo {year} {2006})},\
  \Eprint {http://arxiv.org/abs/nucl-th/0602038} {arXiv:nucl-th/0602038
  [nucl-th]} \BibitemShut {NoStop}%
\bibitem [{\citenamefont {Sawyer}(1989)}]{Sawyer:1989dp}%
  \BibitemOpen
  \bibfield  {author} {\bibinfo {author} {\bibfnamefont {R.~F.}\ \bibnamefont
  {Sawyer}},\ }\href {\doibase 10.1103/PhysRevD.39.3804} {\bibfield  {journal}
  {\bibinfo  {journal} {Phys. Rev.}\ }\textbf {\bibinfo {volume} {D39}},\
  \bibinfo {pages} {3804} (\bibinfo {year} {1989})}\BibitemShut {NoStop}%
\bibitem [{\citenamefont {Hanauske}\ \emph {et~al.}(2016)\citenamefont
  {Hanauske}, \citenamefont {Takami}, \citenamefont {Bovard}, \citenamefont
  {Rezzolla}, \citenamefont {Font}, \citenamefont {Galeazzi},\ and\
  \citenamefont {Stoecker}}]{Hanauske:2016gia}%
  \BibitemOpen
  \bibfield  {author} {\bibinfo {author} {\bibfnamefont {M.}~\bibnamefont
  {Hanauske}}, \bibinfo {author} {\bibfnamefont {K.}~\bibnamefont {Takami}},
  \bibinfo {author} {\bibfnamefont {L.}~\bibnamefont {Bovard}}, \bibinfo
  {author} {\bibfnamefont {L.}~\bibnamefont {Rezzolla}}, \bibinfo {author}
  {\bibfnamefont {J.~A.}\ \bibnamefont {Font}}, \bibinfo {author}
  {\bibfnamefont {F.}~\bibnamefont {Galeazzi}}, \ and\ \bibinfo {author}
  {\bibfnamefont {H.}~\bibnamefont {Stoecker}},\ }\href@noop {} {\  (\bibinfo
  {year} {2016})},\ \Eprint {http://arxiv.org/abs/1611.07152} {arXiv:1611.07152
  [gr-qc]} \BibitemShut {NoStop}%
\bibitem [{\citenamefont {{Takami}}\ \emph {et~al.}(2015)\citenamefont
  {{Takami}}, \citenamefont {{Rezzolla}},\ and\ \citenamefont
  {{Baiotti}}}]{Takami2015}%
  \BibitemOpen
  \bibfield  {author} {\bibinfo {author} {\bibfnamefont {K.}~\bibnamefont
  {{Takami}}}, \bibinfo {author} {\bibfnamefont {L.}~\bibnamefont
  {{Rezzolla}}}, \ and\ \bibinfo {author} {\bibfnamefont {L.}~\bibnamefont
  {{Baiotti}}},\ }\href {\doibase 10.1103/PhysRevD.91.064001} {\bibfield
  {journal} {\bibinfo  {journal} {Phys. Rev. D}\ }\textbf {\bibinfo {volume}
  {91}},\ \bibinfo {eid} {064001} (\bibinfo {year} {2015})},\ \Eprint
  {http://arxiv.org/abs/1412.3240} {arXiv:1412.3240 [gr-qc]} \BibitemShut
  {NoStop}%
\bibitem [{\citenamefont {{Rezzolla}}\ and\ \citenamefont
  {{Zanotti}}(2013)}]{Rezzolla_book:2013}%
  \BibitemOpen
  \bibfield  {author} {\bibinfo {author} {\bibfnamefont {L.}~\bibnamefont
  {{Rezzolla}}}\ and\ \bibinfo {author} {\bibfnamefont {O.}~\bibnamefont
  {{Zanotti}}},\ }\href {\doibase 10.1093/acprof:oso/9780198528906.001.0001}
  {\emph {\bibinfo {title} {Relativistic Hydrodynamics}}}\ (\bibinfo
  {publisher} {Oxford University Press},\ \bibinfo {address} {Oxford, UK},\
  \bibinfo {year} {2013})\BibitemShut {NoStop}%
\bibitem [{\citenamefont {{Bovard}}\ and\ \citenamefont
  {{Rezzolla}}(2017)}]{Bovard2016}%
  \BibitemOpen
  \bibfield  {author} {\bibinfo {author} {\bibfnamefont {L.}~\bibnamefont
  {{Bovard}}}\ and\ \bibinfo {author} {\bibfnamefont {L.}~\bibnamefont
  {{Rezzolla}}},\ }\href@noop {} {\bibfield  {journal} {\bibinfo  {journal}
  {ArXiv e-prints}\ } (\bibinfo {year} {2017})},\ \Eprint
  {http://arxiv.org/abs/1705.07882} {arXiv:1705.07882 [gr-qc]} \BibitemShut
  {NoStop}%
\bibitem [{\citenamefont {{Maione}}\ \emph {et~al.}(2017)\citenamefont
  {{Maione}}, \citenamefont {{De Pietri}}, \citenamefont {{Feo}},\ and\
  \citenamefont {{L{\"o}ffler}}}]{maione2017}%
  \BibitemOpen
  \bibfield  {author} {\bibinfo {author} {\bibfnamefont {F.}~\bibnamefont
  {{Maione}}}, \bibinfo {author} {\bibfnamefont {R.}~\bibnamefont {{De
  Pietri}}}, \bibinfo {author} {\bibfnamefont {A.}~\bibnamefont {{Feo}}}, \
  and\ \bibinfo {author} {\bibfnamefont {F.}~\bibnamefont {{L{\"o}ffler}}},\
  }\href@noop {} {\bibfield  {journal} {\bibinfo  {journal} {arXiv:1707.03368}\
  } (\bibinfo {year} {2017})},\ \Eprint {http://arxiv.org/abs/1707.03368}
  {arXiv:1707.03368 [gr-qc]} \BibitemShut {NoStop}%
\bibitem [{\citenamefont {{Clark}}\ \emph {et~al.}(2016)\citenamefont
  {{Clark}}, \citenamefont {{Bauswein}}, \citenamefont {{Stergioulas}},\ and\
  \citenamefont {{Shoemaker}}}]{Clark2016}%
  \BibitemOpen
  \bibfield  {author} {\bibinfo {author} {\bibfnamefont {J.~A.}\ \bibnamefont
  {{Clark}}}, \bibinfo {author} {\bibfnamefont {A.}~\bibnamefont {{Bauswein}}},
  \bibinfo {author} {\bibfnamefont {N.}~\bibnamefont {{Stergioulas}}}, \ and\
  \bibinfo {author} {\bibfnamefont {D.}~\bibnamefont {{Shoemaker}}},\ }\href
  {\doibase 10.1088/0264-9381/33/8/085003} {\bibfield  {journal} {\bibinfo
  {journal} {Class. Quantum Grav.}\ }\textbf {\bibinfo {volume} {33}},\
  \bibinfo {eid} {085003} (\bibinfo {year} {2016})},\ \Eprint
  {http://arxiv.org/abs/1509.08522} {arXiv:1509.08522 [astro-ph.HE]}
  \BibitemShut {NoStop}%
\bibitem [{\citenamefont {{Bose}}\ \emph {et~al.}(2017)\citenamefont {{Bose}},
  \citenamefont {{Chakravarti}}, \citenamefont {{Rezzolla}}, \citenamefont
  {{Sathyaprakash}},\ and\ \citenamefont {{Takami}}}]{Bose2017}%
  \BibitemOpen
  \bibfield  {author} {\bibinfo {author} {\bibfnamefont {S.}~\bibnamefont
  {{Bose}}}, \bibinfo {author} {\bibfnamefont {K.}~\bibnamefont
  {{Chakravarti}}}, \bibinfo {author} {\bibfnamefont {L.}~\bibnamefont
  {{Rezzolla}}}, \bibinfo {author} {\bibfnamefont {B.~S.}\ \bibnamefont
  {{Sathyaprakash}}}, \ and\ \bibinfo {author} {\bibfnamefont {K.}~\bibnamefont
  {{Takami}}},\ }\href@noop {} {\bibfield  {journal} {\bibinfo  {journal}
  {arXiv:1705.10850}\ } (\bibinfo {year} {2017})},\ \Eprint
  {http://arxiv.org/abs/1705.10850} {arXiv:1705.10850 [gr-qc]} \BibitemShut
  {NoStop}%
\bibitem [{\citenamefont {Sekiguchi}\ \emph {et~al.}(2011)\citenamefont
  {Sekiguchi}, \citenamefont {Kiuchi}, \citenamefont {Kyutoku},\ and\
  \citenamefont {Shibata}}]{Sekiguchi:2011zd}%
  \BibitemOpen
  \bibfield  {author} {\bibinfo {author} {\bibfnamefont {Y.}~\bibnamefont
  {Sekiguchi}}, \bibinfo {author} {\bibfnamefont {K.}~\bibnamefont {Kiuchi}},
  \bibinfo {author} {\bibfnamefont {K.}~\bibnamefont {Kyutoku}}, \ and\
  \bibinfo {author} {\bibfnamefont {M.}~\bibnamefont {Shibata}},\ }\href
  {\doibase 10.1103/PhysRevLett.107.051102} {\bibfield  {journal} {\bibinfo
  {journal} {Phys. Rev. Lett.}\ }\textbf {\bibinfo {volume} {107}},\ \bibinfo
  {pages} {051102} (\bibinfo {year} {2011})},\ \Eprint
  {http://arxiv.org/abs/1105.2125} {arXiv:1105.2125 [gr-qc]} \BibitemShut
  {NoStop}%
\bibitem [{\citenamefont {Paschalidis}\ \emph {et~al.}(2012)\citenamefont
  {Paschalidis}, \citenamefont {Etienne},\ and\ \citenamefont
  {Shapiro}}]{Paschalidis:2012ff}%
  \BibitemOpen
  \bibfield  {author} {\bibinfo {author} {\bibfnamefont {V.}~\bibnamefont
  {Paschalidis}}, \bibinfo {author} {\bibfnamefont {Z.~B.}\ \bibnamefont
  {Etienne}}, \ and\ \bibinfo {author} {\bibfnamefont {S.~L.}\ \bibnamefont
  {Shapiro}},\ }\href {\doibase 10.1103/PhysRevD.86.064032} {\bibfield
  {journal} {\bibinfo  {journal} {Phys. Rev.}\ }\textbf {\bibinfo {volume}
  {D86}},\ \bibinfo {pages} {064032} (\bibinfo {year} {2012})},\ \Eprint
  {http://arxiv.org/abs/1208.5487} {arXiv:1208.5487 [astro-ph.HE]} \BibitemShut
  {NoStop}%
\bibitem [{\citenamefont {{Kaplan}}\ \emph {et~al.}(2014)\citenamefont
  {{Kaplan}}, \citenamefont {{Ott}}, \citenamefont {{O'Connor}}, \citenamefont
  {{Kiuchi}}, \citenamefont {{Roberts}},\ and\ \citenamefont
  {{Duez}}}]{Kaplan2013}%
  \BibitemOpen
  \bibfield  {author} {\bibinfo {author} {\bibfnamefont {J.~D.}\ \bibnamefont
  {{Kaplan}}}, \bibinfo {author} {\bibfnamefont {C.~D.}\ \bibnamefont {{Ott}}},
  \bibinfo {author} {\bibfnamefont {E.~P.}\ \bibnamefont {{O'Connor}}},
  \bibinfo {author} {\bibfnamefont {K.}~\bibnamefont {{Kiuchi}}}, \bibinfo
  {author} {\bibfnamefont {L.}~\bibnamefont {{Roberts}}}, \ and\ \bibinfo
  {author} {\bibfnamefont {M.}~\bibnamefont {{Duez}}},\ }\href {\doibase
  10.1088/0004-637X/790/1/19} {\bibfield  {journal} {\bibinfo  {journal}
  {Astrophys. J.}\ }\textbf {\bibinfo {volume} {790}},\ \bibinfo {eid} {19}
  (\bibinfo {year} {2014})},\ \Eprint {http://arxiv.org/abs/1306.4034}
  {arXiv:1306.4034 [astro-ph.HE]} \BibitemShut {NoStop}%
\bibitem [{\citenamefont {Reisenegger}\ and\ \citenamefont
  {Bonacic}(2003)}]{Reisenegger:2003pd}%
  \BibitemOpen
  \bibfield  {author} {\bibinfo {author} {\bibfnamefont {A.}~\bibnamefont
  {Reisenegger}}\ and\ \bibinfo {author} {\bibfnamefont {A.~A.}\ \bibnamefont
  {Bonacic}},\ }\href@noop {} {\  (\bibinfo {year} {2003})},\ \Eprint
  {http://arxiv.org/abs/astro-ph/0303454} {arXiv:astro-ph/0303454} \BibitemShut
  {NoStop}%
\bibitem [{\citenamefont {Bonacic}(2003)}]{Bonacic:2003th}%
  \BibitemOpen
  \bibfield  {author} {\bibinfo {author} {\bibfnamefont {A.~A.}\ \bibnamefont
  {Bonacic}},\ }\href@noop {} {\bibfield  {journal} {\bibinfo  {journal}
  {Master thesis, Universidad Catolica de Chile}\ } (\bibinfo {year}
  {2003})}\BibitemShut {NoStop}%
\bibitem [{\citenamefont {Alford}\ \emph {et~al.}(2012)\citenamefont {Alford},
  \citenamefont {Reddy},\ and\ \citenamefont {Schwenzer}}]{Alford:2011df}%
  \BibitemOpen
  \bibfield  {author} {\bibinfo {author} {\bibfnamefont {M.~G.}\ \bibnamefont
  {Alford}}, \bibinfo {author} {\bibfnamefont {S.}~\bibnamefont {Reddy}}, \
  and\ \bibinfo {author} {\bibfnamefont {K.}~\bibnamefont {Schwenzer}},\ }\href
  {\doibase 10.1103/PhysRevLett.108.111102} {\bibfield  {journal} {\bibinfo
  {journal} {Phys. Rev. Lett.}\ }\textbf {\bibinfo {volume} {108}},\ \bibinfo
  {pages} {111102} (\bibinfo {year} {2012})},\ \Eprint
  {http://arxiv.org/abs/1110.6213} {arXiv:1110.6213 [nucl-th]} \BibitemShut
  {NoStop}%
\bibitem [{\citenamefont {Alford}\ and\ \citenamefont
  {Pangeni}(2017)}]{Alford:2016cee}%
  \BibitemOpen
  \bibfield  {author} {\bibinfo {author} {\bibfnamefont {M.~G.}\ \bibnamefont
  {Alford}}\ and\ \bibinfo {author} {\bibfnamefont {K.}~\bibnamefont
  {Pangeni}},\ }\href {\doibase 10.1103/PhysRevC.95.015802} {\bibfield
  {journal} {\bibinfo  {journal} {Phys. Rev.}\ }\textbf {\bibinfo {volume}
  {C95}},\ \bibinfo {pages} {015802} (\bibinfo {year} {2017})},\ \Eprint
  {http://arxiv.org/abs/1610.08617} {arXiv:1610.08617 [nucl-th]} \BibitemShut
  {NoStop}%
\bibitem [{\citenamefont {Alford}\ \emph {et~al.}(2015)\citenamefont {Alford},
  \citenamefont {Han},\ and\ \citenamefont {Schwenzer}}]{Alford:2014jha}%
  \BibitemOpen
  \bibfield  {author} {\bibinfo {author} {\bibfnamefont {M.~G.}\ \bibnamefont
  {Alford}}, \bibinfo {author} {\bibfnamefont {S.}~\bibnamefont {Han}}, \ and\
  \bibinfo {author} {\bibfnamefont {K.}~\bibnamefont {Schwenzer}},\ }\href
  {\doibase 10.1103/PhysRevC.91.055804} {\bibfield  {journal} {\bibinfo
  {journal} {Phys. Rev.}\ }\textbf {\bibinfo {volume} {C91}},\ \bibinfo {pages}
  {055804} (\bibinfo {year} {2015})},\ \Eprint {http://arxiv.org/abs/1404.5279}
  {arXiv:1404.5279 [astro-ph.SR]} \BibitemShut {NoStop}%
\bibitem [{\citenamefont {{Radice}}(2017)}]{Radice2017}%
  \BibitemOpen
  \bibfield  {author} {\bibinfo {author} {\bibfnamefont {D.}~\bibnamefont
  {{Radice}}},\ }\href {\doibase 10.3847/2041-8213/aa6483} {\bibfield
  {journal} {\bibinfo  {journal} {Astrophys. J. Lett.}\ }\textbf {\bibinfo
  {volume} {838}},\ \bibinfo {eid} {L2} (\bibinfo {year} {2017})},\ \Eprint
  {http://arxiv.org/abs/1703.02046} {arXiv:1703.02046 [astro-ph.HE]}
  \BibitemShut {NoStop}%
\end{thebibliography}%

\end{document}